\documentclass[12pt,a4paper]{article}

\usepackage{amssymb}
\usepackage{amsmath}
\usepackage{amsfonts}
\usepackage{authblk}
\usepackage{setspace}
\usepackage{graphicx}

\numberwithin{equation}{section}

\newcommand{\BbbR}{\mathbb{R}}

\newcommand{\identoper}{\mathbb{I}}

\newcommand{\mralpha}{{{}_{\text{o}}\alpha}}
\newcommand{\mrbeta}{{{}_{\text{o}}\beta}}

\DeclareMathOperator{\Realpart}{Re}

\DeclareMathOperator{\Imagpart}{Im}

\title{Superconducting circuit boundary conditions\\
beyond the Dynamical Casimir Effect}

\author{Jason Doukas}
\author{Jorma Louko}
\affil{School of Mathematical Sciences, 
University of Nottingham,\\ 
Nottingham NG7 2RD, 
UK}

\date{{\small Revised January 2015. Published in Phys.\ Rev.\ D {\bf 91}, 044010 (2015)}}

\begin{document}

\maketitle
\begin{abstract}
We study analytically the time-dependent boundary 
conditions of superconducting microwave circuit experiments 
in the high plasma frequency limit, in which the conditions are 
Robin-type and relate the value of the field to 
the spatial derivative of the field. 
We give an explicit solution to the field evolution 
for boundary condition modulations that are small in magnitude but may have 
arbitrary time dependence, in a formalism that applies both 
to a semiopen waveguide 
and to a closed waveguide with two independently 
adjustable boundaries. 
The correspondence between the microwave 
Robin boundary conditions and the mechanically-moving 
Dirichlet boundary conditions of the Dynamical Casimir Effect 
is shown to break down at high field frequencies, 
approximately one order of magnitude above the frequencies probed in the 2011 
experiment of Wilson et~al. 
Our results bound the parameter regime in which a 
microwave circuit can be used to model relativistic effects 
in a mechanically-moving cavity, and they show that beyond this 
parameter regime moving mirrors produce more particles 
and generate more entanglement than their non-moving 
microwave waveguide simulations. 
\end{abstract}


\section{Introduction}

The quantum theory of relativistic fields with moving boundaries 
was first explored
by Moore in a remarkably original paper on the quantum formulation of
linearly polarized light in a one-dimensional moving
cavity~\cite{Moore}. The primary result of this investigation was the
discovery that moving mirrors in vacuum create
photons.  Later, motivated by developments in quantum field theory in
curved spacetimes, the specialization to a single moving mirror in
Minkowski spacetime 
was carried out by Fulling and Davies~\cite{Fulling1976},
who again found that non-uniformly accelerating mirrors generate
radiation. These effects, in which particles are produced by moving
boundaries, are generally referred to as the Dynamical Casimir Effect
(DCE) or the 
Nonstationary Casimir Effect \cite{Dodonov:2001yb,Dodonov:2010zza,Dalvit:2011yua}. 

An experimental verification of the DCE in a system with a mechanically moving 
boundary has remained elusive because of the technological
challenges in creating sufficiently large
accelerations~\cite{Dodonov:2001yb,Dodonov:2010zza,Dalvit:2011yua}. 
However, an
experimental observation of a similar particle creation effect has
been reported in a mechanically static semiconductor waveguide where
the boundary condition on the field is modulated electrically, by
a superconducting quantum-interference device
(SQUID)~\cite{Wilson2011}. A~related experimental observation 
of particle creation in Josephson metamaterial 
has been reported in~\cite{Lahteenmaki:2013mda}. 
These observations open fascinating
prospects for simulating on a mechanically static desktop device
quantum phenomena due to motion, including entanglement generation and
degradation, in a regime where the moving system would experience
significant relativistic effects
\cite{friis-tele2012,lindkvist-twin2013,lindkvist-precision2014}.

In this paper we address the evolution of a quantum field in
waveguides of the type used in the experiment of \cite{Wilson2011} in
situations where the modulation of the SQUID(s) at the end(s) of the
waveguide is small in magnitude but may have arbitrary
time-dependence, under the further assumption that the plasma
frequency of the SQUID(s) is negligibly high compared with the
frequencies where the experiment operates. More precisely, recall that the field
$\Phi(t,x)$ in the experiment of \cite{Wilson2011} satisfies, under the
approximations described in~\cite{Johansson:2009zz,Johansson2010},  
the $(1+1)$-dimensional Klein-Gordon
equation 
\begin{align}
0 = \left(\frac{\partial^2}{\partial t^2}  
- v^2 \frac{\partial^2}{\partial x^2}\right)\Phi(t,x) 
\label{eq:waveguide-wave-equation}
\end{align}
with the boundary condition 
\begin{align}
0 = C_J \ddot \phi(t,0) + \left(\frac{2\pi}{\Phi_0}\right)^2
E_J(t) \Phi(t,0)
+ \frac{1}{L_0} \Phi'(t,0)
\ , 
\label{eq:squid-ddot-boundary}
\end{align}
where the cavity is at $x\le0$, the SQUID is at $x=0$, the meaning
of the positive constants $v$, $\Phi_0$, $C_J$ and $L_0$ is as
described in~\cite{Johansson2010}, and $E_J(t)$ can be given arbitrary
time-dependence by modulating the magnetic field applied to the
SQUID\null. In the regime where the SQUID's plasma frequency is large
compared to the frequency of~$\Phi$, the time derivative term in 
\eqref{eq:squid-ddot-boundary} is negligible, and
\eqref{eq:squid-ddot-boundary} reduces to
\begin{align}
0 =  \Phi(t,0) +  L_0^{\text{eff}}(t) \Phi'(t,0)
\ , 
\label{eq:squid-nodot-boundary}
\end{align}
where 
$
L_0^{\text{eff}}(t) = 
{\bigl[\Phi_0/(2\pi)\bigr]}^2 
{\bigl[L_0 E_J(t)\bigr]}^{-1}
$. 
We shall consider the regime in which the boundary condition 
\eqref{eq:squid-nodot-boundary} 
applies and the time-dependence of $L_0^{\text{eff}}(t)$ 
is arbitrary in profile but small in magnitude. 
The regime where the 
SQUID's plasma frequency is large but not 
negligibly so 
is considered in \cite{Wustmann2013,Fosco:2013wpa,Felicetti:2014}. 

Our interest in the boundary condition \eqref{eq:squid-nodot-boundary} is twofold. 
First, considering the condition in its own right, we give 
an explicit solution to the quantum dynamics to leading 
perturbative order in the time variation of the boundary condition, 
using a formalism that allows us to handle both 
the semiopen waveguide 
and a closed waveguide, 
and allowing for the closed waveguide the 
modulations at the two ends to remain independent of each other. 
The mathematical observation underlying our formalism 
is that \eqref{eq:squid-nodot-boundary} 
is a linear relation between the value of the field and the spatial 
derivative of the field, known as a Robin boundary condition, 
and when this condition 
is independent of time, it has a well-known role in 
making the spatial part of the wave operator in 
\eqref{eq:waveguide-wave-equation} self-adjoint 
and hence ensuring the unitarity of the time 
evolution~\cite{reed-simonII,blabk,Bonneau:1999zq}. 
A~time-dependent boundary condition can then be treated perturbatively 
by combining the spectral methods of \cite{reed-simonII,blabk,Bonneau:1999zq} 
to the techniques 
developed in \cite{Bruschi:2011ug,Bruschi:2012pd} 
for mechanically moving cavities. 
For the semiopen waveguide our results agree with 
those found in
\cite{Silva:2011fq,Farina:2012qd,Rego:2013bba,Rego:2014wta} 
via a different formalism. 

Second, we wish to address the sense in which 
\eqref{eq:squid-nodot-boundary} models a 
mechanically moving boundary of the DCE\null. 
For a Dirichlet mirror at the time-dependent location $x = x_{\text{DCE}}(t)$, 
the boundary condition on the field reads 
\begin{align}
0 =  \Phi\bigl(t,x_{\text{DCE}}(t)\bigr) 
\ . 
\label{eq:squid-DCE-boundary}
\end{align}
If we choose in \eqref{eq:squid-nodot-boundary} 
$L_0^{\text{eff}}(t) = x_{\text{DCE}}(t)$, 
\eqref{eq:squid-nodot-boundary} reduces to \eqref{eq:squid-DCE-boundary}
for field frequencies much smaller than $v/|L_0^{\text{eff}}|$, 
but for higher field frequencies the correspondence no longer holds. 
We shall see that our perturbative solution of the field
evolution with the condition \eqref{eq:squid-nodot-boundary} indeed 
differs from the similar perturbative solution 
with the condition \eqref{eq:squid-DCE-boundary}
for frequencies that are not much smaller than $v/|L_0^{\text{eff}}|$, 
both for a semiopen waveguide and a closed waveguide; in particular, 
the large frequency falloff properties of the solution are qualitatively different. 
Simulations of relativistic motion with the mechanically static 
semiconductor waveguide 
would hence need to take place 
in the low frequency regime where the successive 
approximations from \eqref{eq:squid-ddot-boundary} 
via \eqref{eq:squid-nodot-boundary} to \eqref{eq:squid-DCE-boundary} hold. 
Both the experiment of \cite{Wilson2011} 
and the proposals of \cite{friis-tele2012,lindkvist-twin2013,lindkvist-precision2014} 
appear to operate within in this domain by a margin of 
approximately one order of magnitude. 

The plan of the paper is as follows: 

Section \ref{sec:half-open} considers evolution under a 
small discontinuous change in the boundary condition \eqref{eq:squid-nodot-boundary} 
for a semiopen waveguide, and Section 
\ref{sec:closed} presents the similar analysis for the closed waveguide. 
Evolution under small changes in the boundary condition with 
arbitrary time-dependence for both types of waveguides is written 
out in Section~\ref{sec:cont-evolution}. 
Section \ref{sec:comparison} compares the evolution to that under 
Dirichlet boundary conditions at one or two mechanically moving boundaries. 
Section 
\ref{sec:conclusions} presents a summary and concluding remarks. 

Appendix \ref{app:bogoidentities} collects technical identities in a  
perturbative expansion of Bogoliubov coefficients. 
Appendix \ref{app:accboundary} treats the evolution under 
the Dirichlet boundary condition at one 
mechanically moving boundary in a small acceleration 
expansion in which velocities and travel distances are unrestricted, 
adapting to one moving boundary the treatment of mechanically 
rigid cavities given in~\cite{Bruschi:2011ug,Bruschi:2012pd}. 
In appendix \ref{app:entanglement} we derive the first order formula for the 
negativity measure of entanglement in the case when the modes 
have a continuous spectrum.  

\section{Semiopen waveguide\label{sec:half-open}}

In this section we discuss a semiopen waveguide 
under a sudden change in the Robin boundary 
condition~\eqref{eq:squid-nodot-boundary}. 
Subsection \ref{subsec:half-open-static} establishes the notation and 
reviews the known properties of a time-independent Robin boundary condition. 
The sudden change is implemented in subsection~\ref{subsec:half-open-junction}, 
and small sudden changes about the Robin boundary condition 
relevant for the experiment of \cite{Wilson2011}
and about the Dirichlet 
boundary condition are discussed respectively in subsections 
\ref{subsec:half-open-far-from-Dir}
and~\ref{subsec:half-open-near-Dir}.

\subsection{Static boundary condition\label{subsec:half-open-static}}

We adopt units in which the phase velocity $v$ in 
\eqref{eq:waveguide-wave-equation} is set to unity. 
We may hence think of the field as a real scalar field $\phi$ 
on a $(1+1)$-dimensional Minkowski spacetime, 
with the global Minkowski coordinates $(t,x)$ 
and the metric $ds^2 = -dt^2 + dx^2$. 
We take the boundary to be at $x=0$ and the field to 
live in the half $0 \le x <\infty$. 

We consider the massive Klein-Gordon field equation 
\begin{align}
(\partial_t^2 - \partial_x^2 + \mu^2)\phi=0
\ , 
\label{eq:KG-massive}
\end{align}
where $\mu\ge0$ is the mass. The massless special case $\mu=0$ 
reduces to \eqref{eq:waveguide-wave-equation}. 
We keep here $\mu$ general, 
in part because setting $\mu=0$ does not significantly simplify the analysis, 
but also in part in view of prospective future comparisons 
with mechanically moving cavities in situations where transverse dimensions 
may generate a positive $\mu$ 
by Fourier decomposition \cite{Bruschi:2011ug,Bruschi:2012pd,Friis:2013eva}. 

We introduce at $x=0$ the Robin family of boundary conditions
\begin{align}
0 &= 
\phi(t,0) + D \phi'(t,0)
\ ,
\label{eq:halfspace-boundarycondition}
\end{align}
where $D\in\BbbR \cup \{\infty\}$ is a constant independent of~$t$.
The special case $D=0$ gives the Dirichlet boundary condition,
$\phi(t,0)=0$, the special case $D=\infty$ gives the Neumann boundary
condition, $\phi'(t,0)=0$, and all other values of $D$ mix $\phi(t,0)$
and $\phi'(t,0)$.

Any choice for $D$ makes $-\partial_x^2$ self-adjoint
\cite{reed-simonII,blabk,Bonneau:1999zq} and would hence yield 
a consistent quantum theory of a non-relativistic particle on the half-line, 
although the choice $D=0$ 
may in that context be considered less fine-tuned than 
the others~\cite{walton}. 
In the present context of a relativistic quantum field theory, 
we consider only the values of $D$ that make the spectrum of 
$-\partial_x^2 + \mu^2$ strictly positive. 
For $\mu=0$ this means $D \in (-\infty,0] \cup \{\infty\}$, 
and for $\mu>0$ it means 
$D \in (-\infty,0] \cup \{\infty\} \cup (\mu^{-1}, \infty)$. 
A~negative eigenvalue of $-\partial_x^2 + \mu^2$ 
would give a tachyonic instability, and the zero eigenvalue 
that occurs when $\mu>0$ and $D=\mu^{-1}$ would give a zero mode and hence a 
theory without a Fock vacuum. 

The field is quantised in the usual fashion. 
A continuum of mode solutions that are positive frequency 
with respect to $\partial_t$ are
\begin{align}
\phi_k(t,x) = \frac{1}{\sqrt{\pi \omega}} \, e^{-i\omega t} \sin(k x + \delta)
\ ,
\label{eq:half-phimode-k}
\end{align}
where 
$k>0$, $\omega = \sqrt{k^2+\mu^2}$, $\delta$ 
is determined as a function of $k$ from 
\begin{align}
\tan\delta = -kD
\ , 
\end{align}
and we may fix the phase by choosing 
$|\delta| < \pi/2$ for $D\in\BbbR$ 
and $\delta=\pi/2$ for $D=\infty$. 
When $\mu>0$ and 
$\mu^{-1} < D < \infty$, 
there is in addition a discrete ground state mode, given by 
\begin{align}
\phi_g(t,x) 
:= \frac{1}{{\bigl(\mu^2 D^2 -1 \bigr)}^{1/4}} \, e^{-i\sqrt{\mu^2 - D^{-2}} \, t} 
\, e^{-x/D}
\ . 
\label{eq:half-phimode-ground}
\end{align}
Writing the Klein-Gordon inner product 
in the conventions of \cite{birrell-davies} as 
\begin{align}
(\phi,\chi) = -i \int_0^{\infty} 
\bigl(\phi \, (\partial_t \overline\chi) - (\partial_t\phi) \,\overline \chi\bigr) \, dx
\ , 
\label{eq:halfline-KG-ip}
\end{align}
where the overline denotes complex conjugation, 
the continuum modes $\phi_k$ are Dirac-orthonormal, 
\begin{align}
(\phi_k,\phi_{k'}) = \delta(k-k')
\ , 
\end{align}
and when the discrete ground state $\phi_g$ \eqref{eq:half-phimode-ground} exists, 
it is normalised and orthogonal to the continuum modes. 
The Fock space is built on the vacuum that 
is annihilated by the annihilation operators 
associated with the continuum modes $\phi_k$ and with the discrete mode $\phi_g$
when the latter exists.

\subsection{Sudden change in the boundary condition\label{subsec:half-open-junction}}

Suppose that for $t<0$ we use the field modes as introduced above and for 
$t>0$ we use a similar set of field modes with $D$ replaced by~$D'$. 
Denoting the new continuum modes by $\tilde\phi_{k'}$ and the new discrete mode by 
$\tilde\phi_{g'}$, 
we may match the two sets of modes at 
$t=0$ in the notation of \cite{birrell-davies} as 
\begin{subequations}
\label{eq:half-matching}
\begin{align}
\tilde\phi_{k'} 
&= \int_0^\infty \bigl( 
\mralpha_{k' k} \phi_k
+ \mrbeta_{k' k} \overline{\phi_k}
\, 
\bigr) 
\, dk
\ 
+ \mralpha_{k' g} \phi_g
+ \mrbeta_{k' g} \overline{\phi_g}
\ , 
\\[1ex]
\tilde\phi_{g'} 
&= \int_0^\infty 
\bigl( 
\mralpha_{g' k} \phi_k
+ \mrbeta_{g' k} \overline{\phi_k}
\, 
\bigr) 
\, dk
\ 
+ \mralpha_{g' g} \phi_g
+ \mrbeta_{g' g} \overline{\phi_g}
\ , 
\end{align}
\end{subequations} 
where we have included the lower left subscript ${}_{\text{o}}$ 
in the Bogoliubov coefficients $\mralpha$ and $\mrbeta$ to 
indicate that the change in the boundary condition is sudden. 

Expressions for the Bogoliubov coefficients can be found by 
taking
inner products of \eqref{eq:half-matching} 
with the untilded modes and their complex conjugates~\cite{birrell-davies}. 
For the continuum-to-continuum Bogoliubov coefficients, we find
\begin{subequations}
\label{eq:half-c-alphas-and-betas}
\begin{align}
\mralpha_{k' k}
& = 
\cos(\delta-\delta')\delta(k-k')
+ \frac{\sin\delta \sin\delta'}{\pi\sqrt{\omega \omega'}}
\left(\frac{1}{D'} - \frac{1}{D}\right)
P\left(\frac{1}{\omega-\omega'}\right)
\ , 
\label{eq:hw-alpha-gen}
\\[1ex]
\mrbeta_{k' k}
& = 
\frac{\sin\delta \sin\delta'}{\pi\sqrt{\omega \omega'}\,(\omega+\omega')}
\left(\frac{1}{D'} - \frac{1}{D}\right)
\ , 
\label{eq:hw-beta-gen}
\end{align}
\end{subequations}
where $\delta$ and $\delta'$ are determined by 
\begin{subequations}
\begin{align}
\tan\delta &= -kD
\ , 
\\
\tan\delta' &= -k'D'
\ , 
\end{align}
\end{subequations}
and $P$ in \eqref{eq:hw-alpha-gen} denotes the principal 
value in the integration over $k$ at $k=k'$. 
In the special cases $D=0$ and $D'=0$ 
the formulas \eqref{eq:half-c-alphas-and-betas} 
are understood in their well-defined limiting sense. 
The continuum-to-discrete and discrete-to-continuum Bogoliubov coefficients 
will not be needed below and we omit the formulas. 

We note that while $\mrbeta_{k'k}$ \eqref{eq:hw-beta-gen} 
is a function for all $k$ and~$k'$, 
$\mralpha_{k'k}$ \eqref{eq:hw-alpha-gen} has distributional support at $k=k'$ 
because of the Dirac delta in the first term and 
the principal value integral in the second term.

\subsection{Small sudden change: far from 
Dirichlet\label{subsec:half-open-far-from-Dir}}

We now consider the case where $D$ and $D'$ 
are negative and close to each other. 
This is the situation relevant for the 
the waveguide experiment of~\cite{Wilson2011}. 
(Recall that in our conventions the field $\phi$ lives at $x\ge0$,  
while the field $\Phi$ in 
\eqref{eq:waveguide-wave-equation}--\eqref{eq:squid-nodot-boundary} 
lives at $x\le0$.) 

We write 
\begin{subequations}
\begin{align}
D &= - \Lambda
\ , 
\\
D' &= -\Lambda(1+\eta)
\ , 
\end{align}
\end{subequations}
where $\Lambda$ is a positive constant of dimension length 
and $\eta$ is small. The mode expansion contains no discrete 
mode for any value of~$\mu$. 

Expanding the Bogoliubov coefficients 
\eqref{eq:half-c-alphas-and-betas} in~$\eta$, we find 
\begin{subequations}
\label{eq:halfline-pert-bogos-far}
\begin{align}
\mralpha_{k' k}
& = 
\left\{
1 - \frac{\eta^2 {(k\Lambda)}^2}{2 \, {\bigl[1 + {(k\Lambda)}^2\bigr]}^2}
\right\} \delta(k-k')
\notag
\\[1ex]
& \hspace{3ex}
+ \frac{\eta \Lambda k k' }{\pi\sqrt{\omega \omega'} \, 
\sqrt{1 + {(k\Lambda)}^2}\ \sqrt{1 + {(k'\Lambda)}^2}}
\left[1 - \frac{\eta {(k'\Lambda)}^2}{1 + {(k'\Lambda)}^2}\right]
P\left(\frac{1}{\omega-\omega'}\right)
+ O\bigl(\eta^3\bigr)
\ , 
\label{eq:halfline-pert-bogoalpha-far}
\\[1ex]
\mrbeta_{k' k}
& = 
\frac{\eta \Lambda k k' }{\pi\sqrt{\omega \omega'}\,(\omega+\omega') 
\sqrt{1 + {(k\Lambda)}^2}\ \sqrt{1 + {(k'\Lambda)}^2}}
\left[1 - \frac{\eta {(k'\Lambda)}^2}{1 + {(k'\Lambda)}^2}\right]
+ O\bigl(\eta^3\bigr)
\ . 
\end{align}
\end{subequations}
As a consistency check, we have verified that the 
expansion \eqref{eq:halfline-pert-bogos-far} is 
consistent with the identities satisfied by the Bogoliubov 
coefficients, collected in Appendix~\ref{app:bogoidentities}, 
in the sense that the linear order identities 
\eqref{eq:bogo-pert-real1} 
hold and the off-diagonal part of the quadratic order identities 
\eqref{eq:bogo-pert-real2} holds. 
We are not aware of reasons to suspect inconsistencies in the diagonal part of
\eqref{eq:bogo-pert-real2} but we have not undertaken 
the distributional analysis to examine this.

\subsection{Small sudden change: near Dirichlet\label{subsec:half-open-near-Dir}}

As a second regime of interest, we perturb around the Dirichlet boundary condition. 
This case is not directly relevant for the experiment of~\cite{Wilson2011}, 
but we shall see in Section \ref{sec:comparison}
that it exhibits close theoretical 
similarity with a mechanically moving boundary. 

We set $D=0$ and assume $D'$ to be close to~$0$. 
If $D'$ is negative, the result may be obtained from 
\eqref{eq:halfline-pert-bogos-far} 
by writing $\eta\Lambda = -b$ and letting $\Lambda\to0$ 
while $b$ remains finite and negative but small: 
then $D' = b <0$, and 
the Bogoliubov coefficients are given by 
\begin{subequations}
\label{eq:halfline-pert-bogos-near-Dir}
\begin{align}
\mralpha_{k' k}
& = 
\left[
1 - \tfrac12 {(k b)}^2
\right] \delta(k-k')
- \frac{b k k' }{\pi\sqrt{\omega \omega'}} \, 
P\left(\frac{1}{\omega-\omega'}\right)
+ O\bigl({b}^3\bigr)
\ , 
\label{eq:halfline-pert-bogoalpha-near-Dir}
\\[1ex]
\mrbeta_{k' k}
& = 
- \frac{b k k' }{\pi\sqrt{\omega \omega'}\,(\omega+\omega')}
+ O\bigl(b^3\bigr)
\ . 
\end{align}
\end{subequations}
If $D'$ is positive, the $t>0$ theory has a tachyonic instability 
(for all positive $D'$ if $\mu=0$ and for 
$0 < D' < \mu^{-1}$ if $\mu>0$)
because of the negative eigenvalue of $-\partial_x^2 + \mu^2$; 
however, the tachyon is nonperturbative in $D'$, 
and we have verified that just ignoring 
the tachyon and proceeding directly from 
\eqref{eq:half-c-alphas-and-betas} leads again to 
\eqref{eq:halfline-pert-bogos-near-Dir}, where now
$D' = b >0$. 

As a consistency check, we have verified that 
\eqref{eq:halfline-pert-bogos-near-Dir} satisfies the 
linear order Bogoliubov identities 
\eqref{eq:bogo-pert-real1} 
and the off-diagonal part of the quadratic order 
Bogoliubov identities~\eqref{eq:bogo-pert-real2}, 
regardless the sign of~$b$.

\section{Closed waveguide\label{sec:closed}}

In this section we adapt the analysis of Section \ref{sec:half-open} 
to a cavity waveguide that is closed at both ends. 
For simplicity, we treat only the massless field, $\mu=0$.

\subsection{Static boundary condition\label{subsec:closed-static}}

We follow the notation of Section \ref{sec:half-open}, setting $\mu=0$. 
We place the cavity at $0 \le x \le L$, where the positive constant 
$L$ is the length of the cavity. 

We write the static Robin boundary conditions at the ends of the cavity as 
\begin{subequations}
\label{eq:boundarycondition}
\begin{align}
0 &= 
\phi(t,0) + D_1 \phi'(t,0)
\ ,
\label{eq:0-boundarycondition}
\\
0 &= 
\phi(t,L) + D_2 \phi'(t,L)
\ , 
\end{align}
\end{subequations}
where $D_1$ and $D_2$ are constants independent of~$t$, 
taking values in $\BbbR \cup \{\infty\}$. 

Any choice for $D_1$ and $D_2$ makes 
$-\partial_x^2$ self-adjoint \cite{reed-simonII,blabk,Bonneau:1999zq}. 
To avoid instabilities and zero modes, we assume initially
$D_1$ and $D_2$ to be such  
that the spectrum of $-\partial_x^2$ is strictly positive. 
We shall however see in subsection \ref{subsec:closed-near-Dir} 
that a perturbative treatment in $D_1$ and $D_2$ remains consistent
even in the presence of a nonperturbative tachyon, 
on a par with what happened for the semiopen waveguide 
in subsection~\ref{subsec:half-open-near-Dir}. 

The field is again quantised in the usual fashion. 
The Klein-Gordon inner product is as in \eqref{eq:halfline-KG-ip} 
but integrated over $x\in[0,L]$. 
The equation for the eigenvalues can be written 
down using~\eqref{eq:boundarycondition}. 

\subsection{Small sudden change: far from 
Dirichlet\label{subsec:closed-far-from-Dir}}

We consider the case where 
$D_1<0$ and $D_2>0$, and there is a sudden 
but small change in their values. This models a waveguide whose 
each end terminates at a SQUID as in the experiment of~\cite{Wilson2011}. 

For $t < 0$, we set $D_1 = - \kappa_1 L$ and 
$D_2 = \kappa_2 L$, where $\kappa_1$ and $\kappa_2$ are positive dimensionless 
constants. With this boundary condition
$-\partial_x^2$ is positive definite.
The mode functions are 
\begin{align}
\phi_q(t,x) =
\sqrt{\frac{(1+\kappa_1^2q^2)(1+\kappa_2^2q^2)}
{qF(q)}}
\, 
e^{-iqt/L}\sin(qx/L + \delta_q),
\label{eq:q-mode}
\end{align}
where 
\begin{align}
F(q) := (1+\kappa_1 + \kappa_1^2q^2)(1+\kappa_2 +
  \kappa_2^2q^2) - \kappa_1\kappa_2
\ , 
\end{align}
$q$ goes over the positive solutions to 
\begin{align}
\cot q = \frac{\kappa_1 \kappa_2 q^2-1}{(\kappa_1+\kappa_2)q}
\ , 
\label{eq:q-eigenvalue-equation}
\end{align}
$\tan\delta_q = \kappa_1q$, and we choose the phase so that 
$0<\delta_q < \pi/2$. 
There is exactly one $q$ in each interval $m\pi < q < (m+1)\pi$, 
$m=0,1,2,\ldots$. The mode functions are normalised to 
$(\phi_q, \phi_q')=\delta_{q q'}$. 

For $t>0$, we set $D_1 = (- \kappa_1 + \eta_1)L$ and $D_2 = 
(\kappa_2 + \eta_2)L$, where $\eta_1$ and $\eta_2$ 
are dimensionless constants, assumed to be small. 
We work perturbatively in $\eta_1$ and~$\eta_2$, 
setting both of them to be proportional 
to a formal expansion parameter $\eta$ which at the end is set to unity. 
The mode functions are proportional to
$e^{-ikt}\sin(k x +\delta)$ where $k$
and $\delta$ are determined from~\eqref{eq:boundarycondition}. 
We label the mode functions by the positive solutions to
\begin{align}
\cot p = 
\frac{\kappa_1 \kappa_2 p^2-1}{(\kappa_1+\kappa_2)p}
\ , 
\label{eq:p-eigenvalue-equation}
\end{align}
such that $k = p/L + O(\eta)$, we denote them by~$\tilde\phi_{p}$, 
and we choose their phase to agree with that of 
\eqref{eq:q-mode} in the zeroth perturbative order. 
We then find that the eigenfrequencies are given by 
\begin{align}
k_p = 
\left(
1 +  \frac{\eta_1 (1 + \kappa_2^2 p^2) - \eta_2(1+\kappa_1^2 p^2)}{F(p)}
\right) \frac{p}{L}
\ \ + O\bigl(\eta^2\bigr)
\ . 
\label{eq:k-far-Dir}
\end{align} 

The Bogoliubov coefficients are now defined by matching the modes at $t=0$ as 
\begin{align}
\tilde\phi_{p} 
= \sum_q \bigl( 
\mralpha_{p q} \phi_q
+ \mrbeta_{p q} \overline{\phi_q}
\, 
\bigr) 
\ . 
\label{eq:cavity-matching}
\end{align}
Proceeding as in \cite{Bruschi:2011ug,Friis:2013eva}, 
we find 
\begin{subequations}
\label{eq:cav-gen-bogos}
\begin{align}
\mralpha_{pp}
& = 1 + O\bigl(\eta^2\bigr)
\ , 
\\[1ex]
\mralpha_{pq}
& =
\frac{\Bigl[\eta_1\sqrt{(1+\kappa_2^2p^2)(1+\kappa_2^2q^2)}
-{(-1)}^{\varphi_p+\varphi_q} \eta_2
\sqrt{(1+\kappa_1^2p^2)(1+\kappa_1^2q^2)}\, \Bigr] \sqrt{pq}}
{(p-q)\sqrt{F(p)F(q)}}
\notag
\\[1ex]
& \hspace{3ex}
+  O\bigl(\eta^2\bigr)
\ \ \ \text{for $p\ne q$}\ , 
\\[1ex]
\mrbeta_{pq}
& =
- \frac{\Bigl[\eta_1\sqrt{(1+\kappa_2^2p^2)(1+\kappa_2^2q^2)}
- {(-1)}^{\varphi_p+\varphi_q} \eta_2
\sqrt{(1+\kappa_1^2p^2)(1+\kappa_1^2q^2)}\, \Bigr] \sqrt{pq}}
{(p+q)\sqrt{F(p)F(q)}}
\notag
\\[1ex]
& \hspace{3ex}
+  O\bigl(\eta^2\bigr)
\ , 
\end{align}
\end{subequations}
where the map $q\mapsto \varphi_q$ labels the 
consecutive solutions to \eqref{eq:q-eigenvalue-equation} by
consecutive integers. 
The expressions for the order $\eta^2$ terms 
in \eqref{eq:k-far-Dir}
and \eqref{eq:cav-gen-bogos} are lengthy and we suppress them here. 

As a consistency check, 
the linear terms in \eqref{eq:cav-gen-bogos} 
satisfy the linear order Bogoliubov identities~\eqref{eq:bogo-pert-real1}. 
We are not aware of reasons to suspect inconsistencies in the 
quadratic order identities \eqref{eq:bogo-pert-real2} 
but examining these would require a nontrivial evaluation 
of the left-hand side in \eqref{eq:bogo-pert-real2} 
and we have not carried out this evaluation.

\subsection{Small sudden change: 
near Dirichlet\label{subsec:closed-near-Dir}}

We consider also a perturbation around 
the Dirichlet boundary condition. 
If $\eta_1<0$ and $\eta_2>0$, the result may be obtained from \eqref{eq:cav-gen-bogos} 
simply by taking the limit 
$\kappa_1\to0$ and $\kappa_2\to0$. 
The $t<0$ mode functions are given by 
\begin{align}
\phi_n(t,x) = \frac{1}{\sqrt{\pi n}} \, e^{-in\pi t/L}\sin(n\pi x/L) \ ,
\label{eq:Dirichlet-cav-standard-modes}
\end{align}
where $n=1,2,\ldots$. Using positive integers to label both the 
$t<0$ mode functions and the $t>0$ mode functions, we find 
that the $t>0$ eigenfrequencies are given by 
\begin{align}
k_m = \left(1+ (\eta_1 - \eta_2)+
{(\eta_1-\eta_2)}^2\right) \frac{\pi m}{L}
\ \ + O\bigl(\eta^3\bigr)\ ,
\label{eq:k-near-Dir}
\end{align} 
where $m=1,2,\ldots$, and the Bogoliubov coefficients are given by 
\begin{subequations}
\label{eq:near-Dir-Bogos}
\begin{align}
\mralpha_{mm}
& = 1 
- \tfrac{1}{6} 
\bigl(\eta_1^2 + \eta_1\eta_2 + \eta_2^2\bigr) \, m^2\pi^2
+ O\bigl(\eta^3\bigr)
\ , 
\\[1ex]
\mralpha_{mn}
& =
\frac{\bigl(\eta_1 - {(-1)}^{m+n} \eta_2\bigr) \sqrt{mn}}{(m-n)}
\notag
\\[1ex]
& \hspace{3ex}
- 
\frac{\bigl(\eta_1 - {(-1)}^{m+n} \eta_2\bigr)
(\eta_1-\eta_2) \, n \sqrt{mn}}{{(m-n)}^2}
+ O\bigl(\eta^3\bigr)
\ \ \ \text{for $m\ne n$}\ , 
\\[1ex]
\mrbeta_{mn}
& =
- \frac{\bigl(\eta_1 - {(-1)}^{m+n} \eta_2\bigr) \sqrt{mn}}{(m+n)}
\notag
\\[1ex]
& \hspace{3ex}
- 
\frac{\bigl(\eta_1 - {(-1)}^{m+n} \eta_2\bigr)
(\eta_1-\eta_2) \, n \sqrt{mn}}{{(m+n)}^2}
+ O\bigl(\eta^3\bigr)
\ , 
\end{align}
\end{subequations}
where we have now displayed also the order $\eta^2$ terms. 
If $\eta_1\ge0$ and/or $\eta_2\le0$, 
the $t>0$ theory may have tachyonic instabilities or zero modes; 
however, both of these are nonperturbative, and we have verified 
that setting $D_1=D_2=0$ for $t<0$, $D_1 = \eta_1 L$ and $D_2 = \eta_2 L$ 
for $t>0$, ignoring any tachyons or zero modes, and working directly from 
\eqref{eq:boundarycondition} and~\eqref{eq:cavity-matching}, 
yields
\eqref{eq:k-near-Dir}
and \eqref{eq:near-Dir-Bogos} regardless the signs of $\eta_1$ and~$\eta_2$. 

As a consistency check, 
the expressions in \eqref{eq:near-Dir-Bogos} 
can be verified to satisfy the perturbative Bogoliubov 
identities~\eqref{eq:bogo-pert-real12}. 
The elements of the matrix square on the left-hand side of 
\eqref{eq:bogo-pert-real1} are given by absolutely 
convergent sums that can be evaluated by residue techniques.

\section{Boundary condition with arbitrary time-dependence\label{sec:cont-evolution}}

When the boundary condition has arbitrary time dependence 
but the variations remain so small in magnitude that first-order 
perturbation theory suffices, 
the evolution of the field can be obtained by composing the 
sudden changes of Sections \ref{sec:half-open} and \ref{sec:closed} 
and passing to the limit~\cite{Bruschi:2012pd}. 
We discuss first the semiopen waveguide and then the closed waveguide.

\subsection{Semiopen waveguide\label{subsec:cont-halfline}}

For the semiopen waveguide of Section~\ref{sec:half-open}, 
we consider a boundary condition of the form \eqref{eq:halfspace-boundarycondition} 
where $D$ may change in time but only within the interval $t_0 \le t \le t_f$. 

\subsubsection{Far from Dirichlet}

Consider first the far-from Dirichlet case. We write 
$D = -\Lambda\bigl(1+\eta(t)\bigr)$, 
where $\Lambda$ is a positive constant 
and the function $\eta(t)$ 
is vanishing outside the interval 
$t_0 \le t \le t_f$ and satisfies $|\eta(t)| \ll 1$. 

At $t<t_0$, we introduce early time basis modes that are as in 
\eqref{eq:half-phimode-k} but with the replacement 
$e^{-i\omega t} \to e^{-i\omega(t-t_0)}$. 
At $t>t_f$, we similarly introduce late 
time basis modes that are as in 
\eqref{eq:half-phimode-k} but with the replacement 
$e^{-i\omega t} \to e^{-i\omega(t-t_f)}$. 
Let $\alpha_{k'k}$ and $\beta_{k'k}$ be the coefficients in the
Bogoliubov transformation from the early time modes to the late time modes. 
Working perturbatively in~$\eta$, 
we may proceed as in~\cite{Bruschi:2012pd}, 
and the outcome can be read off from formulas (6) and (7) therein. 
We find 
\begin{subequations}
\label{eq:half-fardir-tcoeffs}
\begin{align}
\alpha_{k'k} 
&= 
e^{i\omega'(t_f-t_0)} 
\left( \delta(k-k') + \hat A_{k'k} + O\bigl(\eta^2\bigr) \right) 
\ , 
\\[1ex]
\beta_{k'k} 
&= 
e^{i\omega'(t_f-t_0)} \hat B_{k'k} + O\bigl(\eta^2\bigr)
\ , 
\end{align}
\end{subequations}
where 
\begin{subequations}
\label{eq:half-fardir-pcoeffs}
\begin{align}
\hat A_{k'k} &= 
-\frac{i\Lambda k k' }{\pi\sqrt{\omega \omega'} \, 
\sqrt{1 + {(k\Lambda)}^2}\ \sqrt{1 + {(k'\Lambda)}^2}}
\int_{t_0}^{t_f}
e^{-i(\omega' - \omega)(t-t_0)} \, \eta(t) \, dt
\ , 
\\[1ex]
\hat B_{k'k} &= 
\frac{i\Lambda k k' }{\pi\sqrt{\omega \omega'} \, 
\sqrt{1 + {(k\Lambda)}^2}\ \sqrt{1 + {(k'\Lambda)}^2}}
\int_{t_0}^{t_f}
e^{-i(\omega' + \omega)(t-t_0)} \, \eta(t) \, dt
\ . 
\end{align}
\end{subequations}
When $\mu=0$, 
\eqref{eq:half-fardir-tcoeffs}
and 
\eqref{eq:half-fardir-pcoeffs}
reduce to what was found in 
\cite{Silva:2011fq,Farina:2012qd,Rego:2013bba,Rego:2014wta} 
via a different formalism.

\subsubsection{Near Dirichlet}

In the near-Dirichlet case, we take 
$D = b(t)$, where the function $b(t)$ 
is vanishing outside the interval 
$t_0 \le t \le t_f$. Proceeding as above, we find 
\begin{subequations}
\label{eq:half-neardir-tcoeffs}
\begin{align}
\alpha_{k'k} 
&= 
e^{i\omega'(t_f-t_0)} 
\left( \delta(k-k') + \hat A_{k'k} + O\bigl(b^2\bigr) \right) 
\ , 
\\[1ex]
\beta_{k'k} 
&= 
e^{i\omega'(t_f-t_0)} \hat B_{k'k} + O\bigl(b^2\bigr)
\ , 
\end{align}
\end{subequations}
where 
\begin{subequations}
\label{eq:half-neardir-pcoeffs}
\begin{align}
\hat A_{k'k} &= 
\frac{i k k' }{\pi\sqrt{\omega \omega'}}
\int_{t_0}^{t_f}
e^{-i(\omega' - \omega)(t-t_0)} \, b(t) \, dt
\ , 
\\[1ex]
\hat B_{k'k} &= 
- \frac{i k k' }{\pi\sqrt{\omega \omega'}}
\int_{t_0}^{t_f}
e^{-i(\omega' + \omega)(t-t_0)} \, b(t) \, dt
\ . 
\end{align}
\end{subequations}

\subsection{Closed waveguide\label{subsec:cont-cavity}}

For the closed waveguide of Section~\ref{sec:closed}, 
we consider a boundary condition of the form 
\eqref{eq:boundarycondition} where $D_1$ and $D_2$ 
may change in time 
but only within the interval $t_0 \le t \le t_f$. 
We may proceed as above. 
The only new aspect is that the method of 
\cite{Bruschi:2012pd} needs to be generalised 
to accommodate the linear 
term that appears in the frequencies 
\eqref{eq:k-far-Dir}
and~\eqref{eq:k-near-Dir}. 

\subsubsection{Far from Dirichlet}

In the far-from Dirichlet case, we write 
$D_1 = \bigl(- \kappa_1 + \eta_1(t)\bigr)L$ and $D_2 = 
\bigl(\kappa_2 + \eta_2(t)\bigr)L$, where $\kappa_1$ and $\kappa_2$ 
are positive constants and the functions $\eta_1(t)$ and $\eta_2(t)$
are vanishing outside the interval 
$t_0 \le t \le t_f$ and satisfy $|\eta_1(t)| \ll 1$ 
and $|\eta_2(t)| \ll 1$. 
Indexing the mode functions 
in the notation of subsection~\eqref{subsec:closed-far-from-Dir}, 
writing 
\begin{align}
\omega_p = p/L
\ , \ \ \ 
\omega_q = q/L
\ , 
\end{align}
and proceeding as above, 
we find that the coefficients in the Bogoliubov
transformation from the early time modes to 
the late time modes are given by 
\begin{subequations}
\label{eq:far-dir-cav-bogocontfull}
\begin{align}
\alpha_{pq} 
&= 
e^{i \omega_p(t_f-t_0)} 
\left( \delta_{pq} + \hat A_{pq} + O\bigl(\eta^2\bigr) \right) 
\ , 
\\[1ex]
\beta_{pq} 
&= 
e^{i \omega_p(t_f-t_0)} 
\hat B_{pq} + O\bigl(\eta^2\bigr)
\ , 
\end{align}
\end{subequations}
where 
\begin{subequations}
\label{eq:far-dir-cav-bogocontlin}
\begin{align}
\hat A_{pq} &= 
\frac{i\sqrt{pq}}
{L\sqrt{F(p)F(q)}}
\left[ \sqrt{(1+\kappa_2^2p^2)(1+\kappa_2^2q^2)} 
\int_{t_0}^{t_f} e^{-i(\omega_p - \omega_q)(t-t_0)} \, \eta_1(t) \, dt
\right.
\notag
\\[1ex]
& 
\hspace{7ex}
\left.
- {(-1)}^{\varphi_p+\varphi_q}
\sqrt{(1+\kappa_1^2p^2)(1+\kappa_1^2q^2)}
\int_{t_0}^{t_f} e^{-i(\omega_p - \omega_q)(t-t_0)} \, \eta_2(t) \, dt
\right]
\ , 
\\[1ex]
\hat B_{pq} &= 
- \frac{i\sqrt{pq}}
{L\sqrt{F(p)F(q)}}
\left[ \sqrt{(1+\kappa_2^2p^2)(1+\kappa_2^2q^2)} 
\int_{t_0}^{t_f} e^{-i(\omega_p + \omega_q)(t-t_0)} \, \eta_1(t) \, dt
\right.
\notag
\\[1ex]
& 
\hspace{7ex}
\left.
- {(-1)}^{\varphi_p+\varphi_q}
\sqrt{(1+\kappa_1^2p^2)(1+\kappa_1^2q^2)}
\int_{t_0}^{t_f} e^{-i(\omega_p + \omega_q)(t-t_0)} \, \eta_2(t) \, dt
\right]
\ . 
\end{align}
\end{subequations}

\subsubsection{Near Dirichlet}

In the near-Dirichlet case, we write 
$D_1 = \eta_1(t)L$ and $D_2 = \eta_2(t)L$.
Indexing the past and future mode functions by positive integers
in the notation of subsection~\eqref{subsec:closed-near-Dir}, 
and writing 
\begin{align}
\omega_m = \pi m/L
\ , \ \ \ 
\omega_n = \pi n/L
\ ,  
\label{eq:near-dir-cav-omegas}
\end{align}
we find 
\begin{subequations}
\label{eq:near-dir-cav-bogocontfull}
\begin{align}
\alpha_{mn} 
&= 
e^{i \omega_m(t_f-t_0)} 
\left( \delta_{mn} + \hat A_{mn} + O\bigl(\eta^2\bigr) \right) 
\ , 
\\[1ex]
\beta_{mn} 
&= 
e^{i \omega_m(t_f-t_0)} 
\hat B_{mn} + O\bigl(\eta^2\bigr)
\ , 
\end{align}
\end{subequations}
where 
\begin{subequations}
\label{eq:near-dir-cav-bogocontlin}
\begin{align}
\hat A_{mn} &= 
\frac{i\pi \sqrt{mn}}{L}
\int_{t_0}^{t_f} e^{-i(\omega_m - \omega_n)(t-t_0)} \, 
\bigl(\eta_1(t) - {(-1)}^{m+n} \eta_2(t) \bigr)
\, dt
\ , 
\\[1ex]
\hat B_{mn} &= 
- \frac{i\pi \sqrt{mn}}{L}
\int_{t_0}^{t_f} e^{-i(\omega_m + \omega_n)(t-t_0)} \, 
\bigl(\eta_1(t) - {(-1)}^{m+n} \eta_2(t) \bigr)
\, dt
\ . 
\end{align}
\end{subequations}

\section{Comparison\label{sec:comparison}}

We are now ready to compare the evolution under the 
time-dependent Robin boundary condition to the evolution under the 
Dirichlet condition at a mechanically moving boundary.

\subsection{Semiopen waveguide}

For the semiopen waveguide, the evolution of a field with mass $\mu\ge0$ 
under the time-dependent 
Robin boundary condition
was given in subsection~\ref{subsec:cont-halfline}. 
The evolution of a massless field under a 
Dirichlet condition at a mechanically moving 
boundary is given in Appendix \ref{app:accboundary} 
in terms of the acceleration of the boundary, 
in a small acceleration approximation that 
allows the velocity and the travel distance to remain arbitrary and 
overlaps with the  
DCE literature results \cite{Dodonov:2001yb,Dodonov:2010zza,Dalvit:2011yua} 
in the common domain of validity. 

Comparing 
\eqref{eq:half-neardir-tcoeffs}--\eqref{eq:half-neardir-pcoeffs}
and 
\eqref{eq:half-rel-cont-bogos}--\eqref{eq:half-rel-cont-bogolins},
we see that 
the massless field with
the mechanically moving boundary
can be simulated to the leading order in perturbation theory 
by the $\mu=0$ near-Dirichlet 
Robin boundary condition provided we choose $b(t)$ so that 
$a(\tau) = \partial_\tau^2 b(\tau)$, where $a(\tau)$ is the proper
acceleration of the boundary as a function of its proper time~$\tau$, 
and the modulation starts and ends so gently that both $b$ and $\dot b$ vanish. 
This is precisely the relation one would have 
expected from the low frequency equivalence between 
the Robin boundary condition 
\eqref{eq:squid-nodot-boundary}
and the mechanically moving 
Dirichlet boundary condition~\eqref{eq:squid-DCE-boundary}. 
The simulation is reliable in the frequency range where the 
first-order perturbation theory results on both the 
Robin side and on the mechanical side remain reliable; 
we shall not attempt to quantify this range more precisely, 
but for given $a(\tau)$ and $b(\tau)$ the range 
is unlikely to include arbitrarily high frequencies. 

Comparing further 
\eqref{eq:half-neardir-tcoeffs}--\eqref{eq:half-neardir-pcoeffs}
and 
\eqref{eq:half-rel-cont-bogos}--\eqref{eq:half-rel-cont-bogolins}
with 
\eqref{eq:half-fardir-tcoeffs}--\eqref{eq:half-fardir-pcoeffs}, 
we see that 
the massless field with
the mechanically moving boundary
can be simulated by the $\mu=0$ far-from-Dirichlet 
Robin boundary condition  
provided we choose $\eta(t)$ so that 
$a(\tau) = - \Lambda \partial_\tau^2 \eta(\tau)$, 
the frequencies are much smaller than~$\Lambda^{-1}$, 
and the modulation starts and ends so gently that both 
$\eta$ and $\dot\eta$ vanish. 
Again, this is precisely the outcome one would have expected from 
\eqref{eq:squid-nodot-boundary}
and~\eqref{eq:squid-DCE-boundary}. 
When the frequencies are not much smaller than~$\Lambda^{-1}$, 
the evolution with the Robin boundary condition 
differs from the evolution with the mechanically moving 
boundary because the square root factors 
in \eqref{eq:half-fardir-pcoeffs} differ from unity. 

Investigating these differences further, recall that the beta
coefficients are directly related to the total photon production
number by:
\begin{equation}
N=\int {|\beta_{k'k}|}^2 \, dk \, dk'
\ .
\end{equation}
Taking $\eta(t)$ to be a sinusoidal function:
\begin{equation}
\label{eqn:eta}
\eta(t)=\varepsilon \sin[\omega_d(t-t_0)]
\ ,
\end{equation}
with positive constant $\varepsilon\ll1$ and driving
frequency~$\omega_d$, the integrals in \eqref{eq:half-fardir-pcoeffs}
can be performed exactly. The dominant part of the photon flux occurs
for frequencies below the driving frequency. We therefore define
the photon flux density, $n(\overline{k})$, as a function of the
reduced frequency $\overline{k}:=k'/\omega_d$, by  
\begin{align}
n(\overline{k}) := \omega_d \int_0^\infty 
{|\beta_{\omega_d \overline{k} \; k}|}^2 \, dk
\ , 
\label{eq:fluxdensity-def}
\end{align}
so that 
\begin{align}
N = \int_0^\infty n(\overline{k}) \, d\overline{k}
\ . 
\end{align}

\begin{figure}[p]
\includegraphics[width=0.5\textwidth]{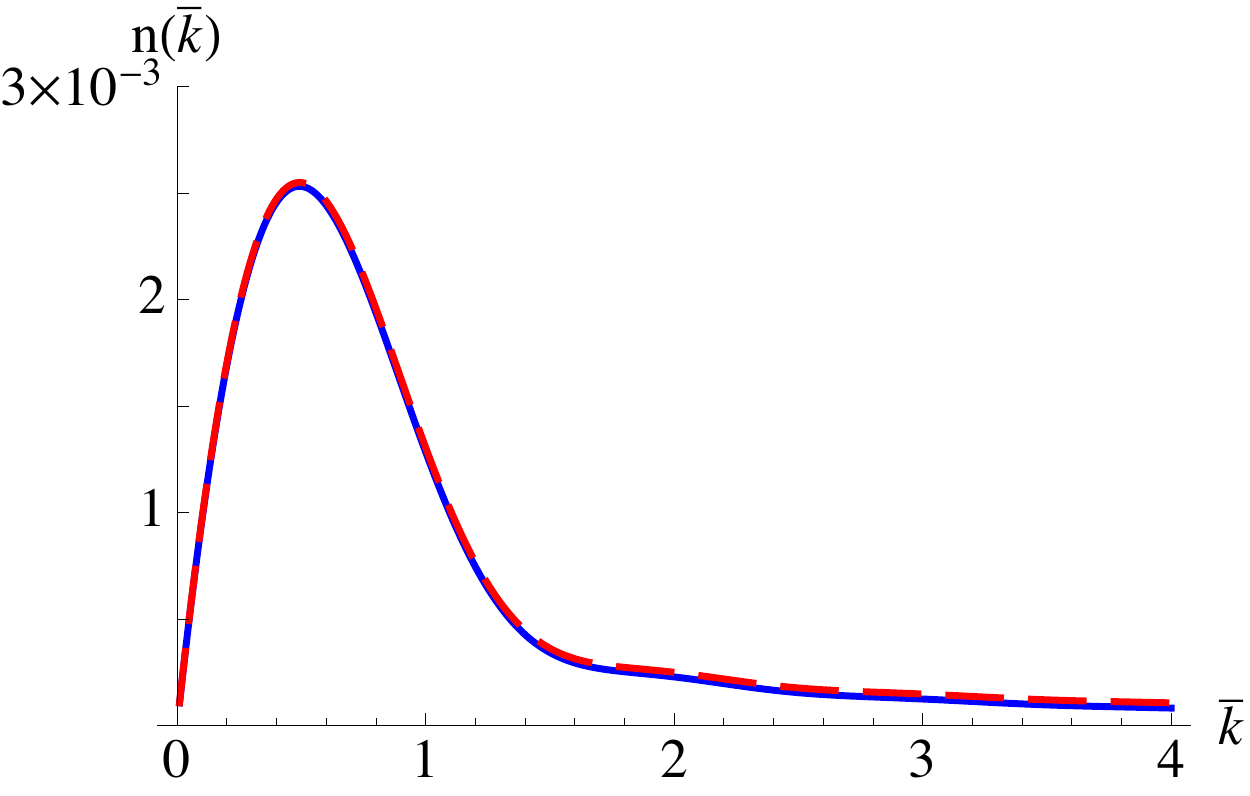}
\includegraphics[width=0.5\textwidth]{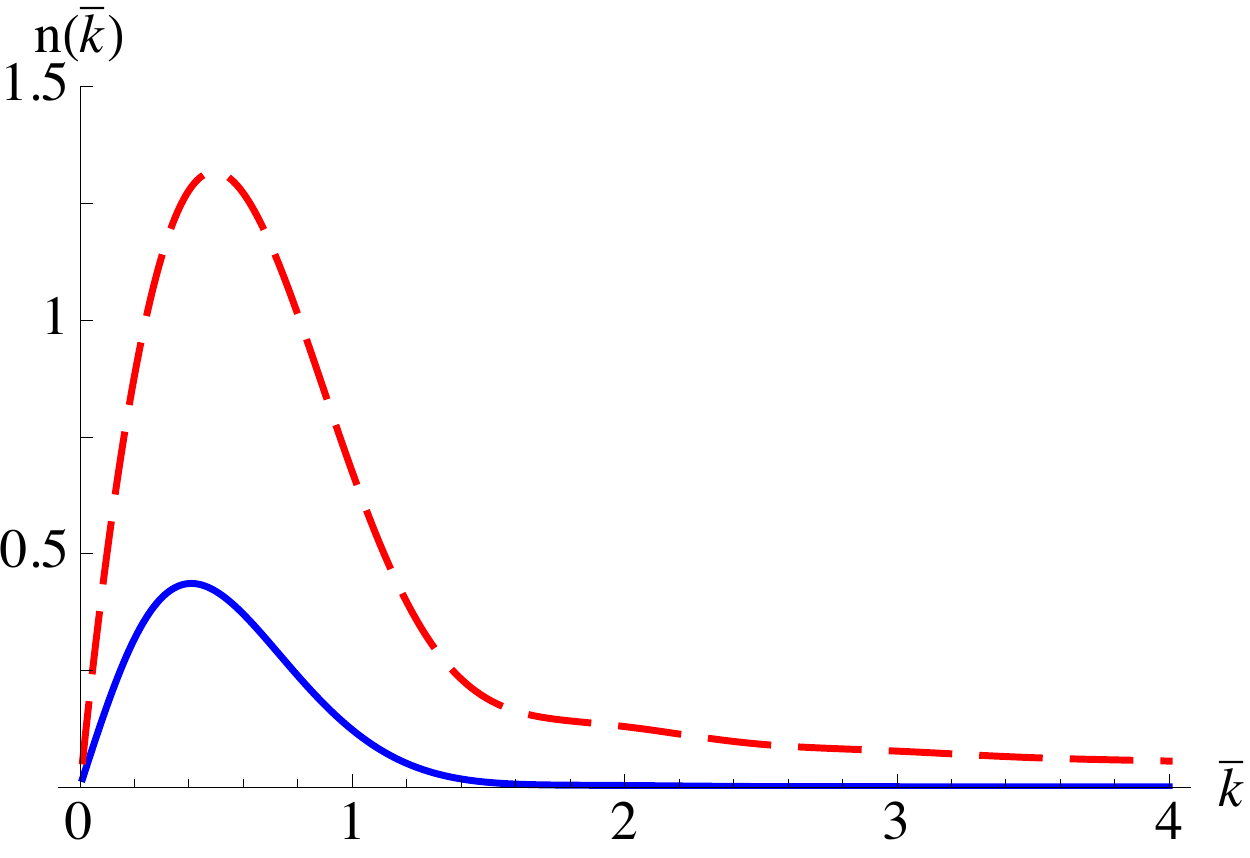}
\caption{\label{fig:photonflux} 
Photon flux densities $n(\overline{k})$ \eqref{eq:fluxdensity-def}
for the semiopen waveguide in terms of the 
reduced frequency $\overline{k}=k/\omega_d$ 
for moving boundary (red dashed) 
and non-moving time-varying Robin boundary (blue solid).  
(Left) In the small effective length limit $\Lambda\ll1/\omega_d$ 
the two systems display near identical behavior. 
(Right) 
In the large effective length limit $\Lambda\gg1/\omega_d$ 
the moving and non-moving systems have different photon production
profiles. 
In both plots we use the values (in units with $v=1$), 
$\epsilon=0.25$, 
$\omega_d=0.155$mm$^{-1}$ and 
$t_f - t_0 =40.5$mm. 
In the Left plot we take $\Lambda=0.44$mm and in the Right 
plot we take $\Lambda=10$mm. 
The parameters in the Left plot can be matched to those 
suggested in \cite{Johansson2010} by taking the 
propagation velocity $v=1.2 \times10^{11}$mm/s and noting that 
$L^\text{eff}_0\rightarrow\Lambda$, 
$\delta L^\text{eff}_0/L^\text{eff}_0\rightarrow\varepsilon$.}
\end{figure}

Figure \ref{fig:photonflux} presents numerical plots of the photon
flux density for $\Lambda\ll1/\omega_d$ and $\Lambda\gg1/\omega_d$. In
both cases the spectrum has in the range $0<\overline{k}<1$ a
distinctive parabolic shape that is qualitatively characteristic of
the DCE, and for $\Lambda\ll1/\omega_d$ 
there is good agreement with the zero temperature curve
of Figure 8 in~\cite{Johansson2010}.

For a quantitative comparison with the DCE, Figure
\ref{fig:photonflux} presents also the photon flux density for the
mechanically moving Dirichlet boundary, obtained from 
\eqref{eq:half-rel-cont-bogos}--\eqref{eq:half-rel-cont-bogolins} 
with the matching $a(\tau) = -
\Lambda \partial_\tau^2 \eta(\tau)$.  When the effective length,
$\Lambda$, is much smaller than the effective wavelength of the
driving frequency ($\Lambda\ll1/\omega_d$), the moving boundary and
Robin-boundary systems produce almost identical spectral functions, 
even though the modulation used for the plot  
has nonvanishing $\dot \eta$ at the start and end. 
On the other hand, when the effective length is larger than the effective
wavelength of the driving frequency ($\Lambda\gg1/\omega_d$), the
spectral functions from the moving boundary system and the
Robin-boundary systems differ: the radiation from the moving boundary
has a higher intensity, and the 
spectrum from the Robin boundary condition
decays more rapidly at high frequencies. Both of these behaviours
result from the square root factors which suppress the beta
coefficients in~\eqref{eq:half-fardir-pcoeffs}.

\begin{figure}[p]
\includegraphics[width=0.5\textwidth]{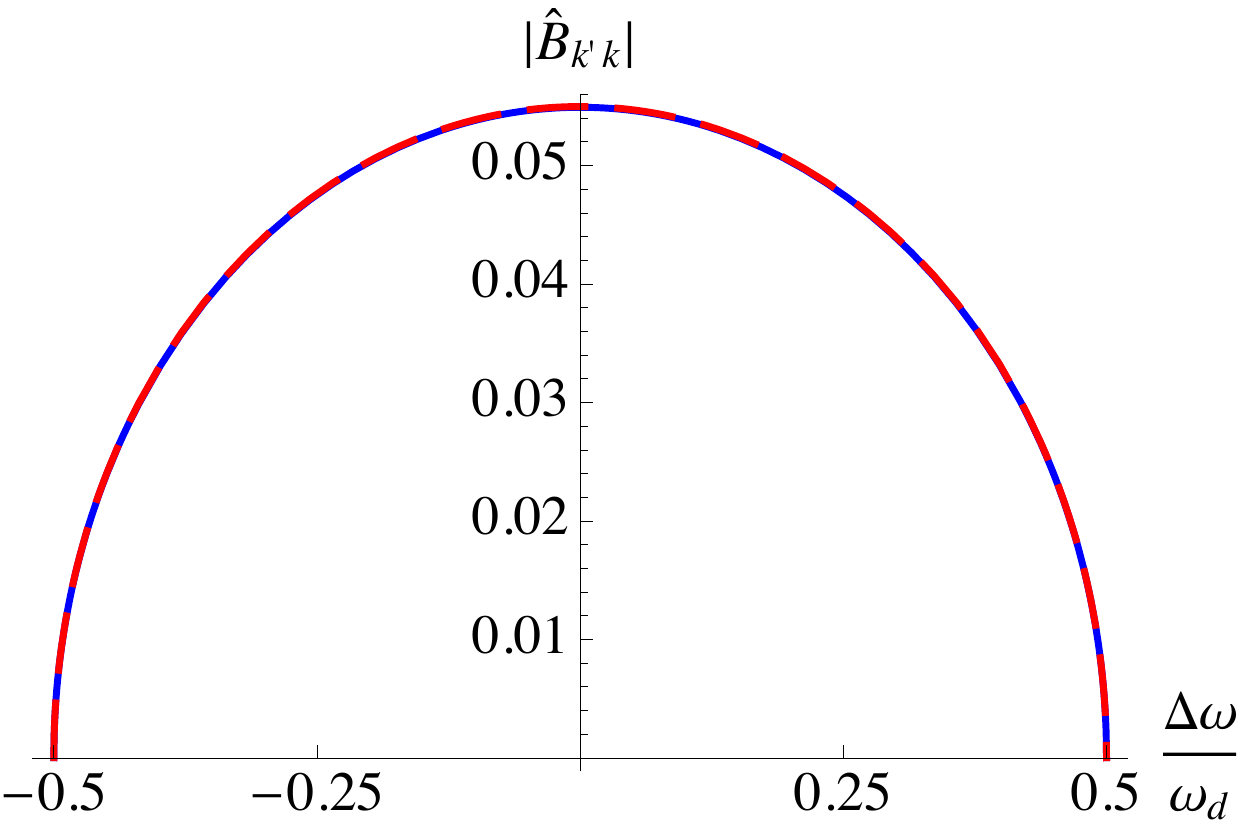}
\includegraphics[width=0.5\textwidth]{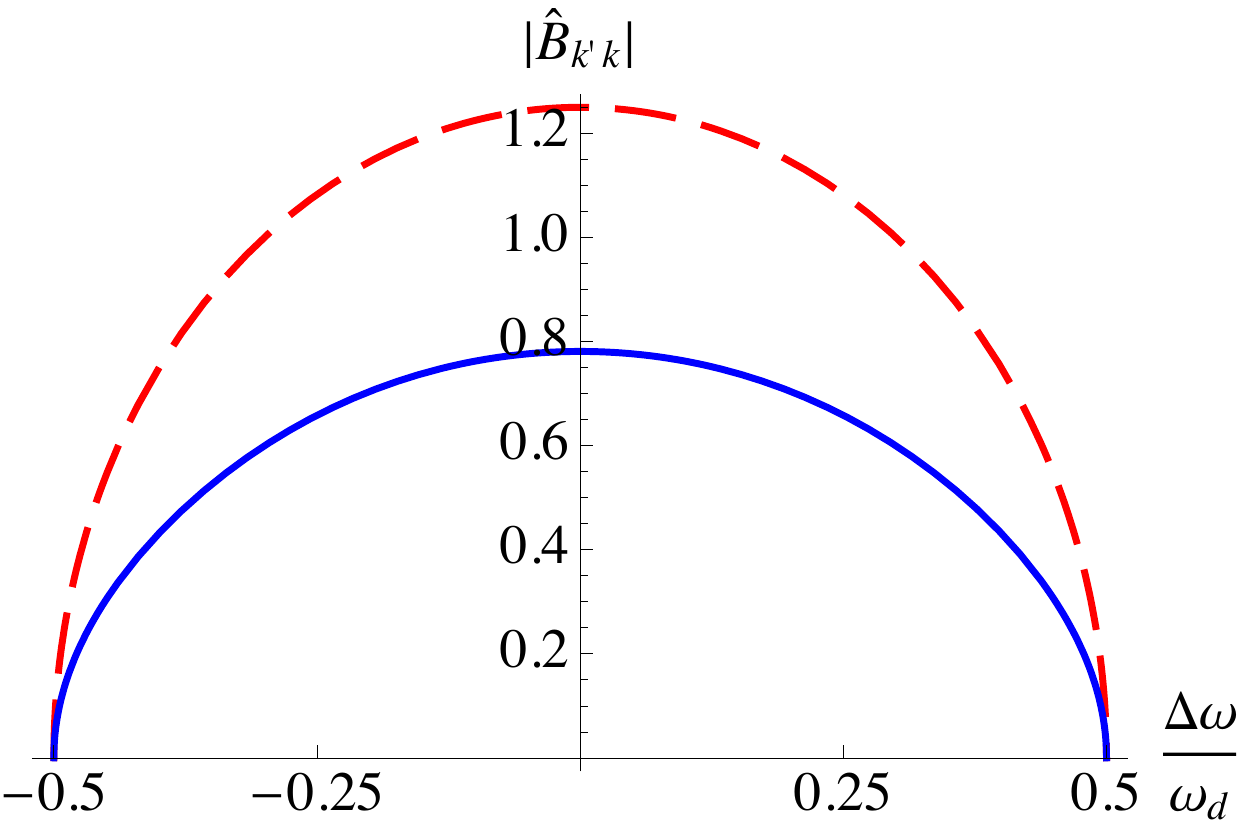}
\caption{\label{fig:negativity} 
The plots show the negativity function $|\hat{B}_{k'k}|
=\mathcal{N}/\Delta k$ (see appendix \ref{app:entanglement}) with 
$k=\omega_d/2+\Delta\omega$ 
and 
$k'=\omega_d/2-\Delta\omega$
in terms of $\Delta\omega/\omega_d$ 
for moving boundary (red dashed) 
and non-moving time-varying Robin boundary (blue solid).  
(Left) In the small effective length limit $\Lambda\ll1/\omega_d$ 
the two systems display near identical behavior. 
(Right) 
In the large effective length limit $\Lambda\gg1/\omega_d$ 
the moving system produces more entanglement than the non-moving system. 
We use the same parameters as in Fig.~\ref{fig:photonflux}.}
\end{figure}

One of the signatures of the DCE radiation is that emitted 
photons are entangled. We can quantify the amount of entanglement 
produced using the negativity \cite{Peres:1996,Horodecki:1996} which is 
defined as minus the sum of the negative eigenvalues of the partially 
transposed state. We show in appendix \ref{app:entanglement} 
that for perturbative Bogoliubov coefficients of the type 
\eqref{eq:half-fardir-tcoeffs}--\eqref{eq:half-fardir-pcoeffs} 
the leading term in the negativity between sharply peaked field 
modes $k$ and $k'$ is given by $\Delta k |\hat{B}_{k'k}|$ 
where $\Delta k$ is the spectral line-width of the frequencies. 

For the driving function given in equation~\eqref{eqn:eta}, 
$|\hat{B}_{k'k}|$ is highest for values of $k+k'$ near the driving frequency. 
Therefore, we define $\Delta\omega=(k-k')/2$, and we expand the frequencies 
near half driving-frequency as:
\begin{subequations}
\begin{align}
k&=\omega_d/2+\Delta\omega
\ ,
\\
k'&=\omega_d/2-\Delta\omega
\ . 
\end{align}
\end{subequations}
The results for $|\hat{B}_{k'k}|$ 
as a function of $\Delta\omega/\omega_d$ are shown 
in Fig.~\ref{fig:negativity}. 
These plots are qualitatively similar to those of 
Fig.~14 in \cite{Johansson2010} 
for the ``without resonator'' line. We again observe good agreement 
when $\Lambda\ll1/\omega_d$ between moving mirrors and non-moving 
microwave waveguides. Also, when $\Lambda\gg1/\omega_d$ we find 
that the moving mirrors generate more entanglement than their 
associated non-moving microwave waveguide simulations. 

The entanglement considered above 
is entanglement between the field modes. This entanglement could be observed directly with a homodyne setup by coupling the emitted radiation into an optics circuit with stationary photodetectors. Another possibility would be to harvest the entanglement by inserting localised detectors, such as an atomic qubit. This has been discussed in a range of settings in \cite{Benenti:2014cta,Benenti:2014pra,Benenti:2014dec,Doukas:2008ju,Doukas:2010wt,Martin-Martinez:2014gra} 
and the references therein.

\subsection{Closed waveguide}

For the closed waveguide, the evolution of a massless field under the
time-dependent Robin boundary condition was given in
subsection~\ref{subsec:cont-cavity}. The evolution of a massless
field in a moving, mechanically rigid cavity of proper length $L$ with
Dirichlet boundary conditions is given by
\cite{Bruschi:2011ug,Bruschi:2012pd}
\begin{subequations}
\label{eq:acc-cav-cont-bogofulls}
\begin{align}
\alpha_{mn} 
&= 
e^{i\omega_m(\tau_f-\tau_0)} 
\left( \delta_{mn} + \hat A_{mn} + O\bigl(a^2\bigr) \right) 
\ , 
\\[1ex]
\beta_{mn} 
&= 
e^{i\omega_m(\tau_f-\tau_0)} \hat B_{mn} + O\bigl(a^2\bigr)
\ , 
\end{align}
\end{subequations}
where 
\begin{subequations}
\label{eq:acc-cav-cont-bogolins}
\begin{align}
\hat A_{nn} &= 0
\ , 
\\[1ex]
\hat A_{mn} &= 
-\frac{i \pi \sqrt{mn} \bigl(1 - {(-1)}^{m+n}\bigr)}{L^2 {(\omega_m-\omega_n)}^2} \, 
\int_{\tau_0}^{\tau_f}
e^{-i(\omega_m - \omega_n)(\tau-\tau_0)} \, a(\tau) \, d\tau
\ \ \ 
\text{for $m\ne n$}
\ , 
\\[1ex]
\hat B_{mn} &= 
\frac{i \pi \sqrt{mn} \bigl(1 - {(-1)}^{m+n}\bigr)}{L^2 {(\omega_m+\omega_n)}^2} \, 
\int_{\tau_0}^{\tau_f}
e^{-i(\omega_m + \omega_n)(\tau-\tau_0)} \, a(\tau) \, d\tau
\ , 
\end{align}
\end{subequations}
$m$ and $n$ are positive integers, $\omega_m$ and $\omega_n$ 
are given by~\eqref{eq:near-dir-cav-omegas}, 
$\tau$ is the proper time at the centre of the cavity 
and $a(\tau)$ is the proper acceleration at the centre of the cavity. 

Comparing 
\eqref{eq:near-dir-cav-bogocontfull}--\eqref{eq:near-dir-cav-bogocontlin}
and 
\eqref{eq:acc-cav-cont-bogofulls}--\eqref{eq:acc-cav-cont-bogolins}, 
we see that 
the massless field in 
the rigidly moving cavity 
can be simulated to the leading order in perturbation theory 
by the near-Dirichlet 
Robin boundary condition provided we choose $\eta_1(\tau) = \eta_2(\tau)$
and $a(\tau) = L \partial_\tau^2 \eta_1(\tau)$, 
and the modulation starts and ends so gently that both $\eta$ and $\dot \eta$ vanish. 
Again, this is precisely the relation one would have 
expected from \eqref{eq:squid-nodot-boundary} and~\eqref{eq:squid-DCE-boundary}, 
and the simulation is reliable in the frequency range where 
the first-order perturbation theory results on both sides are reliable. 

Comparing further with 
\eqref{eq:far-dir-cav-bogocontfull}--\eqref{eq:far-dir-cav-bogocontlin}, 
we see that the rigidly moving cavity can be simulated 
by the far-from-Dirichlet Robin boundary 
condition with $\eta_1(\tau) = \eta_2(\tau)$
and $a(\tau) = L \partial_\tau^2 \eta_1(\tau)$ 
when the frequencies are much smaller than 
$\kappa_1^{-1}$ and~$\kappa_2^{-1}$
and the modulation starts and ends so gently that both 
$\eta$ and $\dot \eta$ vanish. 

We anticipate that the simulation can be extended 
to a moving cavity that is not rigid, 
in the small amplitude regime commonly considered 
in the DCE literature~\cite{Dodonov:2001yb,Dodonov:2010zza,Dalvit:2011yua}, 
by equating $L \eta_1(t)$ (respectively $L \eta_2(t)$) 
to the variation in the position of the left (right) boundary. 
We have however not examined this question systematically.

\section{Conclusions\label{sec:conclusions}}

In this paper we have analysed the evolution of a quantum field in
$1+1$ dimensions under time-dependent Robin boundary conditions that
occur in superconducting microwave circuit experiments in the high
plasma frequency limit.  We solved the evolution explicitly to linear
order in the time variation of the Robin boundary condition, 
in a formalism that allowed us to handle both a 
semiopen waveguide and a closed waveguide, and for the latter allowing 
the boundary conditions at the two ends to be varied
independently. For frequencies that are much smaller than the
effective inverse length associated with the SQUID(s) at the end(s) of
the waveguide, we verified that a suitable modulation of the SQUID(s)
allows the waveguide to simulate a Dirichlet boundary condition at a
mechanically moving boundary of the DCE, even in the regime where the
mechanical motion is relativistic; both the experiment reported in
\cite{Wilson2011} and the experimental proposals of
\cite{friis-tele2012,lindkvist-twin2013,lindkvist-precision2014}
appear to operate within in this domain by a margin of approximately
one order of magnitude.  For higher frequencies the waveguide still
exhibits particle creation and mode mixing, but these can no longer be
quantitatively matched to those of the DCE, and in particular the
large frequency falloff properties in the evolution are qualitatively
different. These features in the large frequency Bogoliubov
coefficients result in differing particle emission spectrum for the
moving and non-moving systems when the effective length is larger than
the inverse driving frequency $L\gg1/\omega_d$. In this limit,
mechanically moving boundary radiation can be characterised as having
a larger total flux and a less steep falloff at high frequencies
compared to radiation from the static waveguide with time-varying
Robin boundary conditions.

While the analogy between a moving Dirichlet boundary and a
time-varying Robin boundary condition is useful for simulation
purposes it is important to keep in mind that the physical systems
corresponding to these two situations are different and therefore can
lead to different outcomes. On the one hand, our results support
proposals to simulate in a mechanically static waveguide quantum
phenomena due to motion, including entanglement generation and
degradation, even in a regime where the mechanically moving system
experiences significant relativistic effects
\cite{friis-tele2012,lindkvist-twin2013,lindkvist-precision2014}.  On
the other hand, our results demonstrate that the interpretation of a
waveguide experiment in terms of the simulation of the DCE is possible
only in certain parameter ranges. Within the Robin boundary condition
approximation that we have analysed, the range of DCE interpretation
is determined just by the effective inverse length scale set by the
SQUID(s) at the end(s) of the waveguide.  It remains a subject to
future work to establish the range of DCE interpretation when all
relevant effects beyond the Robin boundary condition approximation are
considered~\cite{Johansson:2009zz,Johansson2010}, including the
effects due to a large but finite SQUID plasma frequency analysed
in \cite{Wustmann2013,Fosco:2013wpa,Felicetti:2014}.

\section*{Acknowledgments}

J.L. thanks 
Ivette Fuentes, 
Joel Lindkvist 
and 
Carlos Sab\'in 
for helpful discussions and correspondence on waveguides 
and David Bruschi for helpful discussions on 
equations~\eqref{eq:mr-alphabeta-int}. 
We thank 
Giuliano Benenti, 
Simone Felicetti, 
Diego Mazzitelli
and 
Hector Silva
for bringing earlier work to our attention. 
J.D. was supported 
in the early stages of the work by EPSRC 
(Career Acceleration Fellowship 
Grant EP/G00496X/2 to Ivette Fuentes). 
J.L. was supported in part by STFC (Theory
Consolidated Grant ST/J000388/1).

\begin{appendix}

\section{Perturbative Bogoliubov 
identities\label{app:bogoidentities}}

In this appendix we record identities satisfied by a perturbative
expansion of the Bogoliubov coefficients. These identities are used in
the main text for consistency checks of the perturbative treatment.

Let $\alpha$ and $\beta$ denote the matrices of a 
Bogoliubov transformation in the notation of~\cite{birrell-davies}, 
with the matrix elements $\alpha_{jk}$ and~$\beta_{jk}$. 
The indices may be discrete or continuous; 
in the latter case the matrix is understood 
as the kernel of an integral operator and may 
include a distributional part. By construction, the matrices satisfy the 
Bogoliubov identities 
\begin{subequations}
\label{eq:bogo-idents}
\begin{align}
\identoper & = \alpha\alpha^\dagger - \beta\beta^\dagger
\ , 
\\
0 & = \alpha \beta^T - \beta \alpha^T
\ , 
\end{align}
\end{subequations}
where $\identoper$ stands for the identity matrix. 


Suppose now that $\alpha$ and $\beta$ have the formal power series expansions
\begin{subequations}
\label{eq:bogo-mat-expansions}
\begin{align}
\alpha & = 
\identoper + \epsilon \alpha_1 + \epsilon^2 \alpha_2 + \epsilon^3 \alpha_3 + \cdots 
\ , 
\\
\beta & = \epsilon \beta_1 + \epsilon^2 \beta_2 + \epsilon^3 \beta_3 + \cdots 
\ ,   
\end{align}
\end{subequations}
where $\epsilon$ is a real-valued expansion parameter. 
Substituting \eqref{eq:bogo-mat-expansions} in \eqref{eq:bogo-idents} 
and collecting terms order by order,
order $\epsilon^0$ is identically satisfied while orders 
$\epsilon^1$ and $\epsilon^2$ give 
\begin{subequations}
\label{eq:bogo-pert-complex12}
\begin{align}
\epsilon^1: & 
\ \ 
\begin{cases}
0 = \alpha_1 + \alpha_1^\dagger
\ , 
\\
0 = \beta_1 - \beta_1^T
\ , 
\end{cases}
\\[1ex]
\epsilon^2: & 
\ \ 
\begin{cases}
0 = \alpha_1 \alpha_1^\dagger - \beta_1 \beta_1^\dagger + \alpha_2 + \alpha_2^\dagger
\ , 
\\
0 = \beta_1 \alpha_1^T - \alpha_1 \beta_1^T + \beta_2 - \beta_2^T
\ . 
\end{cases}
\end{align}
\end{subequations}
When $\alpha$ and $\beta$ are real, \eqref{eq:bogo-pert-complex12} simplifies to 
\begin{subequations}
\label{eq:bogo-pert-real12}
\begin{align}
\epsilon^1: & 
\ \ 
\begin{cases}
\alpha_1 = - \alpha_1^T
\ , 
\\
\beta_1 = \beta_1^T
\ , 
\end{cases}
\label{eq:bogo-pert-real1}
\\[1ex]
\epsilon^2: & 
\ \ 
{(\alpha_1 \pm \beta_1)}^2
= 
\alpha_2 + \alpha_2^T \pm \bigl( \beta_2 - \beta_2^T \bigr)
\ . 
\label{eq:bogo-pert-real2}
\end{align}
\end{subequations}

\section{Accelerated boundary in Minkowski spacetime\label{app:accboundary}}

In this appendix we consider a quantised massless scalar field in $(1+1)$-dimensional 
Minkowski spacetime subject to the Dirichlet boundary 
condition at one accelerated boundary, 
in the limit where the acceleration is treated perturbatively 
but may have arbitrary time-dependence, and the velocity and travel distance of 
the boundary remain arbitrary. 
We follow the methods that were developed 
for a mechanically rigid accelerated cavity in \cite{Bruschi:2011ug,Bruschi:2012pd}. 
The results overlap with those in the 
DCE literature \cite{Dodonov:2001yb,Dodonov:2010zza,Dalvit:2011yua} 
for small amplitude oscillations 
in the common domain of validity. 
The corresponding problem for a classical scalar field 
has been analysed in 
\cite{dittrich01,dittrich02,dittrich03}.

\subsection{Inertial boundary to uniformly accelerated boundary}

We work with a massless scalar field $\phi$ in $(1+1)$-dimensional 
Minkowski spacetime, in the notation of the main text: 
the metric is $ds^2 = - dt^2 + dx^2$,
and the wave equation is \eqref{eq:KG-massive} with $\mu=0$. 

For $t<0$, we take the field to live in the half-space $x\ge a^{-1}$, 
where $a$ is a positive constant of dimension inverse length, 
and we adopt at $x=a^{-1}$ the Dirichlet boundary condition. 
We adopt the basis of mode functions
\begin{align}
\phi_k(t,x) = \frac{1}{\sqrt{\pi k}} \, e^{-i k t}\sin[k(x-a^{-1})] 
\ , 
\label{eq:r-staticmode}
\end{align}
where~$k>0$. $\phi_k$~has the positive frequency 
$k$ with respect to~$\partial_t$, 
and the normalisation in the Klein-Gordon inner product is  
$(\phi_k,\phi_{k'}) = \delta(k-k')$. 

For $t\ge0$, we make the boundary follow the uniformly-accelerated worldline 
$x = \sqrt{t^2+a^{-2}}$. The proper acceleration of the boundary is equal to~$a$. 
The field lives in the region $x \ge \sqrt{t^2+a^{-2}}$, 
and we take the field to satisfy the Dirichlet boundary condition 
at the accelerated boundary. We adopt the basis of mode functions 
\begin{align}
\Phi_k(t,x) = \frac{1}{2 i \, \sqrt{\pi k}} 
\left\{ 
\bigl[a(x-t)\bigr]^{i k/a}
- 
\bigl[a(x+t)\bigr]^{-i k/a}
\right\}
\ , 
\label{eq:r-rmode}
\end{align}
where $k>0$. 
$\Phi_k$~has the positive frequency 
$k/a$ with respect to 
the boost Killing vector $x\partial_t + t\partial_x$, 
and it has the positive frequency $k$ 
with respect to the proper time of the boundary. 
The normalisation in the Klein-Gordon inner product is  
$(\Phi_k,\Phi_{k'}) = \delta(k-k')$. 

At the junction $t=0$, we match the two 
sets of modes by the Bogoliubov transformation 
\begin{align}
\Phi_{k'} 
&= \int_0^\infty \bigl( 
\mralpha_{k' k} \phi_k
+ \mrbeta_{k' k} \overline{\phi_k}
\, 
\bigr) 
\, dk
\ . 
\end{align}
From the inner products that give the 
Bogoliubov coefficients~\cite{birrell-davies}, we find 
\begin{subequations}
\label{eq:mr-alphabeta-int}
\begin{align}
\mralpha_{k'k}
& = \frac{1}{\pi a}
\sqrt{\frac{k'}{k}}
\, \Realpart \left[
\int_1^\infty \frac{dy}{y} \, y^{-ik'/a} \, e^{i(k/a)(y-1)}
\right]
\ , 
\label{eq:mr-alpha-int}
\\[1ex]
\mrbeta_{k'k}
& = \frac{1}{\pi a}
\sqrt{\frac{k'}{k}}
\, \Realpart \left[
\int_1^\infty \frac{dy}{y} \, y^{ik'/a} \, e^{i(k/a)(y-1)}
\right]
\ . 
\label{eq:mr-beta-int}
\end{align}
\end{subequations}

\subsection{Small acceleration expansion}

We wish to find the asymptotic form of $\mralpha_{k'k}$ and 
$\mrbeta_{k'k}$ \eqref{eq:mr-alphabeta-int}
when the acceleration $a$ of the boundary is small compared 
with both $k$ and~$k'$. 

The small $a$ expansion of $\mrbeta_{k'k}$ 
may be obtained by applying to \eqref{eq:mr-beta-int} the method of 
repeated integration by parts~\cite{wong}. The result is 
\begin{align}
\mrbeta_{kk'}
& = 
\frac{a \, \sqrt{k'k}}{\pi{(k+k')}^3}
+ O\bigl(a^{3}\bigr)
\ . 
\label{eq:mr-beta-exp}
\end{align}

The small $a$ expansion of $\mralpha_{k'k}$
is more involved since 
we expect the coefficients in this expansion 
to be no longer functions but distributions, 
the leading term being $\delta(k-k')$. 
We shall not give a rigorous treatment
but proceed heuristically as follows. 

Starting from the integral in~\eqref{eq:mr-alpha-int}, 
we replace $k$ in the integrand by $k+i\epsilon$ 
where $\epsilon>0$. 
If $\epsilon$ is considered fixed, the asymptotic small $a$ expansion 
can be obtained by the method of 
repeated integration by parts~\cite{wong}, with the result 
\begin{align}
\mralpha_{k'k}
& = \frac{1}{\pi}
\sqrt{\frac{k'}{k}}
\left\{
- \Imagpart\left[
\frac{1}
{{(k-k' + i\epsilon)}}
\right]
+ 
a \Realpart\left[
\frac{k+i\epsilon}
{{(k-k' + i\epsilon)}^3}
\right]
\vphantom{\left[
\frac{(k+i\epsilon) 
\bigl[2(k+i\epsilon)+k'\bigr]}{{(k-k' + i\epsilon)}^4}
\right]}
\right.
\notag
\\[1ex]
& \hspace{12ex}
+
\left.
a^2 
\Imagpart\left[
\frac{(k+i\epsilon) 
\bigl[2(k+i\epsilon)+k'\bigr]}{{(k-k' + i\epsilon)}^5}
\right]
+ O\bigl(a^{3}\bigr)
\right\}
\ .
\label{eq:mr-alpha-2nd-raw}
\end{align}
Each of the three terms shown in \eqref{eq:mr-alpha-2nd-raw} 
has a well-defined distributional limit as $\epsilon\to0$, 
as follows from the identity 
\begin{align}
\lim_{\epsilon\to0^+} 
\int 
dk \, \frac{f(k)}{k + i\epsilon}
= 
-i \pi f(0) + P\int dk \, \frac{f(k)}{k} \ ,
\end{align}
and its derivatives. 
Assuming
that the $\epsilon\to0$ limit commutes with the small $a$ expansion, 
we obtain 
\begin{align}
\mralpha_{k'k}
& = \delta(k-k')
+ 
a
G_1(k', k)
+ 
a^2 G_2(k',k)
+ O\bigl(a^{3}\bigr)
\ , 
\label{eq:mr-alpha-2nd-limit}
\end{align}
where the distributions $G_1$ and $G_2$ are given by 
\begin{subequations}
\begin{align}
\int_0^\infty dk \, G_1(k',k) \, f(k) 
& = 
\frac{{(k')}^{1/2}}{2\pi}
P \int_0^\infty dk
\,\frac{1}{{(k-k')}}
\, \partial_k^2 \left( k^{1/2} \, f(k)\right)  
\ , 
\\[1ex]
\int_0^\infty dk \, G_2(k',k) \, f(k) 
& = 
- \frac{{(k')}^{1/2}}{12}  \, \partial^4_{k'} 
\left( {(k')}^{3/2} f(k')\right)  
- \frac{{(k')}^{3/2}}{24}  \, \partial^4_{k'} 
\left( {(k')}^{1/2} f(k')\right)  
\ . 
\end{align}
\end{subequations}
Both $G_1$ and $G_2$ are hence representable by a 
function except at the coincidence limit, and we may write 
\begin{subequations}
\begin{align}
G_1(k',k) &= \frac{\sqrt{k k'}}{\pi\,{(k-k')}^3} 
\hspace*{-15ex}
&
\text{for $k\ne k'$}
\ , 
\\[1ex]
G_2(k',k) &= 0
\hspace*{-15ex}
&\text{for $k\ne k'$}
\ . 
\end{align}
\end{subequations}

As a consistency check, 
the expansions in \eqref{eq:mr-beta-exp} and \eqref{eq:mr-alpha-2nd-limit} 
satisfy the linear order Bogoliubov identities~\eqref{eq:bogo-pert-real1}. 
The quadratic order Bogoliubov identities 
\eqref{eq:bogo-pert-real2} would require a distributional treatment 
and we shall not analyse them here.

\subsection{Arbitrarily-accelerated boundary}

Let $\tau$ denote the proper time of the boundary and $a(\tau)$ 
the proper acceleration of the boundary, such that positive 
(negative) values of $a(\tau)$ mean 
acceleration towards increasing (decreasing)~$x$. 
We assume that $a(\tau)$ is vanishing outside 
the interval $\tau_0 \le \tau \le \tau_f$
and non-negative within this interval, and we assume 
that $a(\tau)$ remains much smaller than the frequencies to be considered. 

We define the early (respectively late) time modes by
\eqref{eq:r-staticmode} in the inertial frame in which the boundary is
at rest in the early (late) times. Proceeding as in
\cite{Bruschi:2011ug,Bruschi:2012pd}, or in Section
\ref{sec:cont-evolution} of the present paper, we find that the
Bogoliubov coefficients from the early time modes to the late time
modes are
\begin{subequations}
\label{eq:half-rel-cont-bogos}
\begin{align}
\alpha_{k'k} 
&= 
e^{ik'(\tau_f-\tau_0)} 
\left( \delta(k-k') + \hat A_{k'k} + O\bigl(a^2\bigr) \right) 
\ , 
\label{eq:half-rel-cont-bogoalpha}
\\[1ex]
\beta_{k'k} 
&= 
e^{ik'(\tau_f-\tau_0)} \hat B_{k'k} + O\bigl(a^2\bigr)
\ , 
\end{align}
\end{subequations}
where 
\begin{subequations}
\label{eq:half-rel-cont-bogolins}
\begin{align}
\hat A_{k'k} &= 
-\frac{i \sqrt{k k'}}{\pi {(k-k')}^2} \, 
\int_{\tau_0}^{\tau_f}
e^{-i(k' - k)(\tau-\tau_0)} \, a(\tau) \, d\tau
\ \ \ 
\text{for $k\ne k'$}
\ , 
\\[1ex]
\hat B_{k'k} &= 
\frac{i \sqrt{k k'}}{\pi {(k+k')}^2} \, 
\int_{\tau_0}^{\tau_f}
e^{-i(k' + k)(\tau-\tau_0)} \, a(\tau) \, d\tau
\ , 
\end{align}
\end{subequations}
and we omit the examination of the distributional part of 
$\hat A_{k'k}$ at $k=k'$. 

The above treatment 
assumes $a(\tau)$ to be non-negative. 
We shall not examine the validity of \eqref{eq:half-rel-cont-bogolins} 
when $a(\tau)$ may take negative values. 

\section{Entanglement formula for continuous spectra\label{app:entanglement}}

In this appendix we derive the formula for the perturbative
approximation to the bipartite mode entanglement of the field, as
measured by the entanglement negativity, in the case when the field
solutions have continuous spectra. These generalise the negativity
formulas found for field solutions with discrete eigenvalues in
\cite{Friis:2012}. The field is prepared initially in the vacuum state
and is subjected to an evolution which can be described by Bogoliubov
transformations that are assumed to take the general form:
\begin{subequations}
\label{generalevolution}
\begin{align}
\alpha_{k'k} 
&= 
c(k') 
\left( \delta(k-k') + \hat A_{k'k} + O\bigl(\eta^2\bigr) \right) 
\ , 
\\[1ex]
\beta_{k'k} 
&= 
c(k') \hat B_{k'k} + O\bigl(\eta^2\bigr)
\ , 
\end{align}
\end{subequations}
where $\eta\ll1$, $k$ and $k'$ are continuous real-valued parameters,
$\hat A_{k'k}$ and $\hat B_{k'k}$ are of linear order in $\eta$ and
$c(k)$ is a phase taking the general form, $c(k)=e^{if(k)}$, where $f$
is a real-valued function of~$k$. Furthermore, we assume that $\hat
A_{k'k}$ and $\hat B_{k'k}$ satisfy the conditions:
\begin{subequations}
\label{eqn:ABconds-coll}
\begin{eqnarray} \label{eqn:Aantisymm}
\hat A_{k'k}&=&-\hat A_{k k'}^{\star}
\ \ \ 
\text{for $k\ne k'$}, 
\\
\hat B_{k'k}&=&\hat B_{kk'},\label{eqn:Bsymm}
\end{eqnarray} 
\end{subequations}
where the star denotes complex conjugation. 
Equations \eqref{eqn:ABconds-coll}
are satisfied by the
evolution in the semiopen waveguide 
\eqref{eq:half-fardir-tcoeffs} 
and by the evolution with the accelerated 
mirror~\eqref{eq:half-rel-cont-bogolins}. 

The added difficulty of systems with continuous spectra is that the
states are Dirac $\delta$-normalised which can lead to apparent
infinities in formulas for the entanglement if not correctly handled.
The infinities are an artefact of working with infinitely precise
frequencies which in practice is not possible --- there is always a
spectral line-width, $\Delta k$, associated with the measurement of a
frequency. Experiments which measure bipartite entanglement of the
field do so by measuring two distinct frequencies each with a small
line-width. For simplicity, we will assume that the modes measured are
uniform wavepackets of frequencies having spectral line-width $\Delta
k$ and central frequencies $k_\text{p}$, $k_\text{p}'$ respectively.
We define:
\begin{align}
f_{k_\text{p}}(k)=
\begin{cases}
\frac{1}{\sqrt{\Delta k}} 
& \text{for $k\in (k_\text{p}-\frac{\Delta
    k}{2},k_\text{p}+\frac{\Delta k}{2})$}, 
\\[1ex]
0 & \text{otherwise.}
\end{cases}
\end{align}
We will also assume that the measured frequencies are sufficiently
separated such that there is no overlap of their supports in frequency
space:
\begin{equation}
\int f_{k_\text{p}}(k) f_{k_\text{p}'}(k)dk=0. 
\end{equation}

Let $\hat{a}(k)$ be the annihilation operator associated with the
frequency $k$. After the evolution the annihilation operators are
transformed by the Bogoliubov transfomations according to:
\begin{eqnarray}
\hat{\bar{a}}(k)=\int 
\left( 
\alpha_{kk'}^{\star}\hat{a}(k')-\beta_{kk'}^{\star}\hat{a}(k')^{\dagger} 
\right) dk'.
\end{eqnarray}
The quadrature operators associated with these frequencies are:
\begin{subequations}
\begin{eqnarray}
\hat{\bar{x}}(k)&=&\hat{\bar{a}}(k)+\hat{\bar{a}}(k)^{\dagger},\\
\hat{\bar{p}}(k')&=&\frac{1}{i}\left(\hat{\bar{a}}(k)-\hat{\bar{a}}(k)^{\dagger}\right),
\end{eqnarray}
\end{subequations}
where we use the conventions of~\cite{Weedbrook2012}. 

As already alluded these frequencies are not measured precisely rather
the actual measurement occurs over a small band of frequencies. We can
define the quadrature operators associated with a wavepacket centred
at the frequency $k_\text{p}$ by:
\begin{subequations}
\begin{eqnarray}
\hat{\bar{x}}_{k_\text{p}}&=&\int f_{k_\text{p}}(k) \hat{\bar{x}}(k)dk,\\
\hat{\bar{p}}_{k_\text{p}}&=&\int f_{k_\text{p}}(k) \hat{\bar{p}}(k)dk.
\end{eqnarray}
\end{subequations}
A short calculation shows that the quadrature operators satisfy the
commutation relations:
\begin{subequations}
\begin{eqnarray}
\left[\hat{\bar{x}}_{i},\hat{\bar{p}}_{j}\right]&=&2i\delta_{ij},\\
\left[\hat{\bar{x}}_{i},\hat{\bar{x}}_{j}\right]&=& \left[\hat{\bar{p}}_{i},\hat{\bar{p}}_{j}\right]=0,
\end{eqnarray}
\end{subequations}
where $i,j\in\{k_\text{p}, k_\text{p}'\}$.

Since the Bogoliubov transformations are linear and the vacuum state
is a Gaussian state, it follows that the final state of the field will
also be a Gaussian state. Gaussian states are completely characterised
by their first and second statistical moments. It is the second
moments which are important for determining the amount of entanglement
in a Gaussian state. The second moments can be arranged into a
covariance matrix: we define
$R=(\hat{\bar{x}}_{k_\text{p}},\hat{\bar{p}}_{k_\text{p}},\hat{\bar{x}}_{k_\text{p}'},\hat{\bar{p}}_{k_\text{p}'})$, 
then the covariance matrix can be defined by:
\begin{eqnarray}\label{eqn:CM}
\sigma_{ij}=\frac{1}{2} \langle\{R_i ,R_j\}\rangle -\langle R_i\rangle\langle R_j\rangle,
\end{eqnarray}
where curly braces denote anti-commutator and the covariance matrix is
normalised such that the covariance matrix of the vacuum state is the
identity matrix.  Expectation values are to be taken with respect to
the initial state which in our case is the vacuum state. It is easy to
see that the second term in \eqref{eqn:CM} vanishes.

We define the kernels:
\begin{subequations}
\begin{eqnarray}
X(k,k')&=& \frac{1}{2} \langle\{\hat{\bar{x}}(k) ,\hat{\bar{x}}(k')\}\rangle\nonumber\\
&=&\int dl(\alpha_{kl}^{\star}-\beta_{kl})(\alpha_{k'l}-\beta_{k'l}^{\star}),\\
H(k,k')&=& \frac{1}{2} \langle\{\hat{\bar{x}}(k) ,\hat{\bar{p}}(k')\}\rangle\nonumber\\
&=&i\int dl(\alpha_{kl}^{\star}-\beta_{kl})(\alpha_{k'l}+\beta_{k'l}^{\star})-i\delta(k-k'),\\
P(k,k')&=& \frac{1}{2} \langle\{\hat{\bar{p}}(k) ,\hat{\bar{p}}(k')\}\rangle\nonumber\\
&=&\int dl(\alpha_{kl}^{\star}+\beta_{kl})(\alpha_{k'l}+\beta_{k'l}^{\star}),
\end{eqnarray}
\end{subequations}
and the matrix:
\begin{eqnarray}
S(k,k')=\left(\begin{array}{cc}
X(k,k') & H(k,k')\\
H(k',k) & P(k,k')
\end{array}\right),
\end{eqnarray}
then the covariance matrix takes the form:
\begin{eqnarray}
\sigma=\left(\begin{array}{cc}
\sigma_a & \sigma_c\\
\sigma_c^T & \sigma_b
\end{array}\right),
\end{eqnarray}
where superscript $T$ indicates matrix transposition and
\begin{subequations}  
\begin{eqnarray}
\sigma_a&\equiv&\int f_{k_\text{p}}(k)f_{k_\text{p}}(k')S(k,k')dkdk',\\
\sigma_b&\equiv&\int f_{k_\text{p}'}(k)f_{k_\text{p}'}(k')S(k,k')dkdk',\\
\sigma_c&\equiv&\int f_{k_\text{p}}(k)f_{k_\text{p}'}(k')S(k,k')dkdk'.
\end{eqnarray}  
\end{subequations}  

Using relations \eqref{generalevolution} 
and~\eqref{eqn:ABconds-coll}, 
the matrix $S(k,k')$ simplifies to
\begin{eqnarray} \label{eqn:Sexpansion}
S(k,k')=\delta(k-k')+S^1(k,k')+\mathcal{O}(\eta^2),
\end{eqnarray}
where the matrix elements of $S^1(k,k')$ are given by:
\begin{subequations}  
\begin{eqnarray}
S^1_{11}=-S^1_{22}&=&-c(k)^{\star}\beta_{k'k}^{\star}-c(k')\beta_{kk'},\\
S^1_{21}=-S^1_{12}&=&i\big(c(k)^{\star}\beta_{k'k}^{\star}-c(k')\beta_{kk'}\big).
\end{eqnarray}
\end{subequations}  
We can also write the covariance matrix as a Taylor expansion in $\eta$:
\begin{eqnarray}
\sigma=\sigma^0+\sigma^1 +\mathcal{O}(\eta^2),
\end{eqnarray} 
and with equation (\ref{eqn:Sexpansion}) it is easily seen that
$\sigma^0$ is the $4\times4$ identity matrix.

To determine the bipartite entanglement, it is necessary to calculate
the symplectic eigenvalues, $\tilde{\nu}_{\pm}$, of the covariance
matrix of the partially transposed state, $\tilde{\sigma}$. The
partial transpose is implemented by a transformation of the covariance
matrix given by \cite{Simon:2000}:
\begin{subequations}  
\begin{eqnarray}
\sigma_a&\rightarrow&\tilde{\sigma}_a=\sigma_a,\\
\sigma_b&\rightarrow&\tilde{\sigma}_b=\sigma_3\sigma_b\sigma_3,\\
\sigma_c&\rightarrow&\tilde{\sigma}_c=\sigma_c\sigma_3,
\end{eqnarray}
\end{subequations}  
where $\sigma_3=\text{diag}(1,-1)$ is the $z$-direction Pauli matrix.

The symplectic values of $\tilde{\sigma}$ can be found from the
absolute values of the eigenvalues of the matrix
$\Omega\tilde{\sigma}$, where $\Omega=\text{diag}(\omega,\omega)$ and
\begin{eqnarray}
\omega=\left(\begin{array}{cc}
0&1\\
-1&0
\end{array}
\right).
\end{eqnarray}

The zeroth order eigenvalues are $\pm i$ and both eigenvalues have
double degeneracy. It is therefore necessary to use degenerate
perturbation theory to determine the eigenvalues to the linear
order. The linear order corrections to the eigenvalues are found to
be:
\begin{eqnarray}
\lambda_{\pm}=\pm2i \sqrt{I},
\end{eqnarray}
where
\begin{equation}
\label{quadintegral}
I\equiv\int dl\int dl' \int dk \int dk' 
f_{k_\text{p}}(l)f_{k_\text{p}'}(l')f_{k_\text{p}}(k)f_{k_\text{p}'}(k')c(l')c(k)^{\star}\beta_{ll'}\beta_{k'k}^{\star}.
\end{equation}
The symplectic eigenvalues are therefore:
\begin{eqnarray}
\tilde{\nu}_{\pm}=|1\pm2\sqrt{I}|,
\end{eqnarray}
and the negativity is:
\begin{eqnarray}
\mathcal{N}=\text{max}\left(0,\frac{1-\tilde{\nu}_s}{2\tilde{\nu}_s}\right),
\end{eqnarray}
where $\tilde{\nu}_s$ is the smallest of the two symplectic
values. Let us now assume that the wavepackets are very sharply peaked
$\Delta k/k_{\text{p}}\ll1$ and $\Delta k/k_{\text{p}}'\ll1$, then the
integrals in equation \eqref{quadintegral} can be approximated by:
\begin{eqnarray}
I&=&\Delta
k^2c(k_\text{p}')c(k_\text{p})^{\star}\beta_{k_\text{p}k_\text{p}'}\beta_{k_\text{p}'k_\text{p}}^{\star}
\nonumber\\
&=&\Delta k^2|\hat{B}_{k_\text{p}k_\text{p}'}|^2,
\end{eqnarray}
and the negativity (to first order in $\eta$) simplifies to:
\begin{eqnarray}
\mathcal{N}=\Delta k |\hat{B}_{k_\text{p}k_\text{p}'}|.
\label{eq:app:cont-Negformula}
\end{eqnarray}
This is the continuous spectrum formula for the entanglement
negativity and holds for non-integer values of the frequencies. It is
equivalent to the formula for discrete frequency modes in the case
when the modes have opposite parity, i.e., when $(k + k')$ is
odd~\cite{Friis:2012}.

\end{appendix}

\end{document}